\documentclass[reprint,superscriptaddress,prb,aps]{revtex4-2}

\usepackage{mathtools}
\usepackage{amsmath}
\usepackage{amsthm}
\usepackage{amssymb}
\usepackage{graphicx}
\usepackage{bm}
\usepackage{siunitx}
\usepackage[colorlinks,linkcolor=blue, urlcolor=blue,citecolor=blue,bookmarks=false]{hyperref}

\begin{document}

\title{Valley polarization of Landau levels in the ZrSiS surface band driven by residual strain}

\author{Christopher J. Butler}
\email{christopher.butler@riken.jp}
\affiliation{RIKEN Center for Emergent Matter Science, 2-1 Hirosawa, Wako, Saitama 351-0198, Japan}

\author{Masayuki Murase}
\affiliation{Laboratory for Materials and Structures, Tokyo Institute of Technology, Kanagawa 226-8501, Japan}

\author{Shunki Sawada}
\affiliation{Laboratory for Materials and Structures, Tokyo Institute of Technology, Kanagawa 226-8501, Japan}

\author{Ming-Chun Jiang}
\affiliation{RIKEN Center for Emergent Matter Science, 2-1 Hirosawa, Wako, Saitama 351-0198, Japan}
\affiliation{Department of Physics and Center for Theoretical Physics, National Taiwan University, Taipei 10617, Taiwan}

\author{Daisuke Hashizume}
\affiliation{RIKEN Center for Emergent Matter Science, 2-1 Hirosawa, Wako, Saitama 351-0198, Japan}

\author{Guang-Yu Guo}
\affiliation{Department of Physics and Center for Theoretical Physics, National Taiwan University, Taipei 10617, Taiwan}
\affiliation{Physics Division, National Center for Theoretical Sciences, Taipei 10617, Taiwan}

\author{Ryotaro Arita}
\affiliation{RIKEN Center for Emergent Matter Science, 2-1 Hirosawa, Wako, Saitama 351-0198, Japan}
\affiliation{Department of Physics, University of Tokyo, 7-3-1 Hongo Bunkyo-ku, Tokyo 113-0033, Japan}

\author{Tetsuo Hanaguri}
\email{hanaguri@riken.jp}
\affiliation{RIKEN Center for Emergent Matter Science, 2-1 Hirosawa, Wako, Saitama 351-0198, Japan}

\author{Takao Sasagawa}
\affiliation{Laboratory for Materials and Structures, Tokyo Institute of Technology, Kanagawa 226-8501, Japan}
\affiliation{Research Center for Autonomous Systems Materialogy, Tokyo Institute of Technology, Kanagawa 226-8501, Japan}

\begin{abstract}

In a multi-valley electronic band structure, lifting of the valley degeneracy is associated with rotational symmetry breaking in the electronic fluid, and may emerge through spontaneous symmetry breaking order, or through a large response to a small external perturbation such as strain. In this work we use scanning tunneling microscopy to investigate an unexpected rotational symmetry breaking in Landau levels formed in the unusual floating surface band of ZrSiS. We visualize a ubiquitous splitting of Landau levels into valley-polarized sublevels. We demonstrate methods to measure valley-selective Landau level spectroscopy, to infer unknown Landau level indices, and to precisely measure each valley's Berry phase in a way that is agnostic to the band structure and topology of the system. These techniques allow us to measure each valley's low-energy dispersion and infer a rigid valley-dependent contribution to the band energies. Ruling out spontaneous symmetry breaking by establishing the sample-dependence of this valley splitting, we explain the effect in terms of residual strain. A quantitative estimate indicates that uniaxial strain can be measured to a precision of $< \SI{0.025}{\%}$. The extreme valley-polarization of the Landau levels results from as little as $\sim \SI{0.1}{\%}$ strain, and this suggests avenues for manipulation using deliberate strain engineering.

\end{abstract}

\maketitle

\section{Introduction}

Numerous interesting quantum phases and phenomena are associated with the breaking of one or more crystal symmetries by the electronic fluid \cite{Fradkin2010,Coissard2022}. This can apply to the breaking of rotational symmetry, which can be more dramatic in materials whose ground state hosts multiple energy-degenerate band extrema separated by a large crystal momentum. These extrema are referred to as valleys, and a broken-symmetry state allowed by the lifting of the degeneracy is said to be valley-polarized.

An instability towards a rotational symmetry breaking state may result spontaneously from many-body correlations, and examples of the resulting states include recently discovered nematic \cite{Lawler2010,Feldman2016,Bohmer2022,Butler2022} and smectic \cite{Yim2018,Yuan2021,Venkatesan2024} electronic liquid crystals. Many-body correlations can be promoted in intrinsic flatband systems or within Landau levels (LLs) \cite{Fradkin1999,Lilly1999,Feldman2016}. Valley-polarized LLs can host further unusual phenomena such as intrinsic in-plane electric polarization \cite{Sodemann2017, Randeria2018}.

Lifting of the valley degeneracy does not necessarily require many-body correlations however, and can also be caused by external perturbations to the single-particle band structure, such as strain \cite{Feldman2016,Huang2023}. This can lead to ambiguity in the mechanism underlying an observed symmetry breaking state. In the very recent example of a kagom\'{e} lattice superconductor, rotational symmetry breaking that might be attributable to flatband correlations has appeared in some measurements, including scanning tunneling microscopy (STM) measurements \cite{Xiang2021,Nie2022}, but has been absent in others \cite{Liu2023,Frachet2023}, and notably is absent when deliberate steps are taken to eliminate strain \cite{Guo2023}.

Although residual strain is a ubiquitous feature of STM and other measurements, there has been very little study of its typical scale and potential impact, which may be especially relevant for measurements of narrow spectroscopic features of current and future interest such as flatbands and Landau levels, and of sensitive correlated order. Although there exist methods of characterizing relative variations in highly inhomogeneous strain \cite{Lawler2010}, presently there is no way of directly detecting and quantifying homogeneous or weakly varying strain, unless for very large values of about one percent or more. A more in-depth understanding of it will aid efforts to investigate and properly characterize observed symmetry breaking electronic states, and may also help to guide the exploitation of strain to achieve new functionality \cite{Dai2019,Hosoi2024}.

\begin{figure*}
\centering
\includegraphics[scale=1]{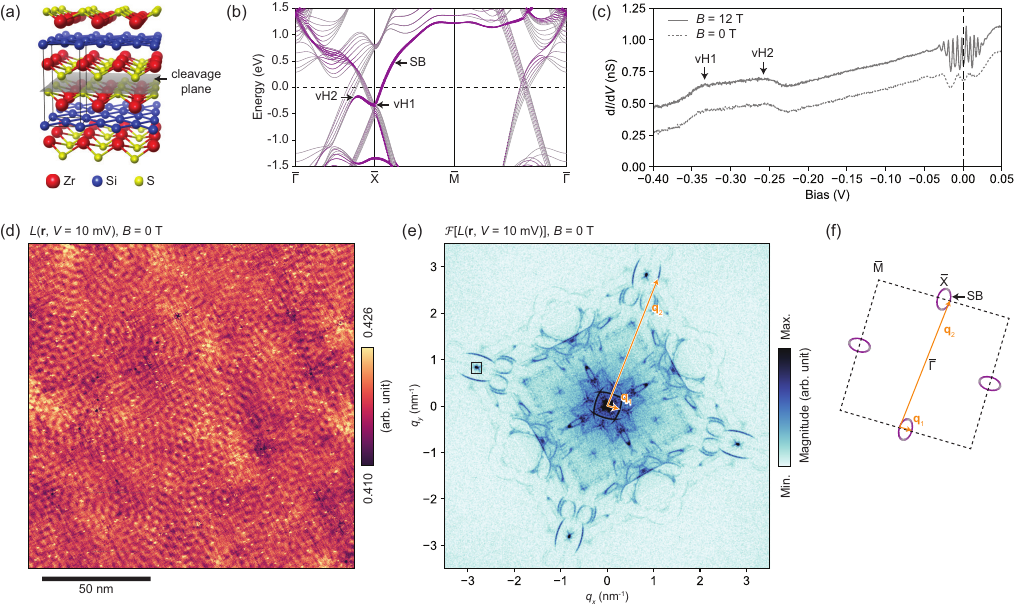}
\caption{\label{fig:1}
Overview of the ZrSiS surface band as seen using STM. (a) Structure of ZrSiS showing a cleavage plane in gray, depicted using VESTA \cite{VESTA}. (b) Calculated surface band structure, with the surface projection of bulk bands in gray, and surface bands in purple. The relative size of the markers represents the relative weight of the wavefunction projected to the surface. The floating surface band is indicated (SB), as well as two nearly-flat parts of the surface band that result in van Hove-like features (vH1 and vH2). (c) Tunneling conductance [$\frac{\mathrm{d}I}{\mathrm{d}V}(V)$] curves acquired under a magnetic field of $B = \SI{12}{T}$ and at zero field, showing the emergence of LLs near $E_{\mathrm{F}}$ ($V = \SI{50}{mV}$, $I_{\mathrm{set}} = \SI{50}{pA}$, $V_{\mathrm{mod.}} = \SI{0.5}{mV}$). The curve for $\SI{12}{T}$ is vertically offset by $\SI{0.2}{nS}$. The features vH1 and vH2 are marked, corresponding to those in (b). (d) Normalized conductance $L \left( \mathbf{r}, V = \SI{10}{mV} \right)$ acquired in a $150 \times \SI{150}{nm^{2}}$ field-of-view ($V = \SI{50}{mV}$, $I_{\mathrm{set}} = \SI{500}{pA}$, $V_{\mathrm{mod.}} = \SI{5}{mV}$), showing modulations due to quasiparticle interference. (e) Fourier transform of the image in (d), symmetrized according to the expected $C_{4v}$ surface symmetry. A reciprocal lattice peak is marked with a black square. Two quasiparticle scattering vectors associated with the surface band are marked as $\mathbf{q}_{1}$ and $\mathbf{q}_{2}$. (f) Depiction of the iso-energetic contour of the surface band near $E_{F}$ (the bulk bands are neglected). The origins of $\mathbf{q}_{1}$ and $\mathbf{q}_{2}$ in scattering within and between surface band pockets, respectively, are shown.}
\end{figure*}

Here we investigate ZrSiS, a material chiefly known for its Dirac nodal-line band structure \cite{Schoop2016}. We focus on its unusual `floating' surface band. As first described by Topp \textit{et al.} \cite{Topp2017}, while non-symmorphic symmetry enforces a band degeneracy at the bulk Brillouin zone boundary, this symmetry breaks down at the surface. The degeneracy is lifted and a surface band is then allowed to split off, and this creates a highly two-dimensional and fairly simple multi-valley system atop the bulk nodal-line semimetal.

Using high-resolution tunneling spectroscopy under high magnetic fields, we show that the LLs formed within the surface band generally exhibit a splitting into pairs of sublevels. Imaging of the quasiparticle interference patterns in each sublevel shows that they are strongly valley-polarized, and allows direct $r$- and $q$-space visualizations of the orientation of the valley-split system. We demonstrate an efficient method for obtaining valley-selective tunneling spectra that has broad applicability to multi-valley systems. We also introduce a model-free method for inferring the indices of observed LLs even where the lowest LL cannot be seen, and this method has the added outcome of a direct measurement of the Berry phase around the LL orbit. These methods allow us to separately and precisely describe the dispersion for each valley of the surface bands. Comparison of the dispersion curves for different pairs of valleys indicates a rigid valley-dependent contribution to the band energies.

The observation of sample dependence of the valley splitting, as well as a case of zero splitting, rules out spontaneous symmetry breaking order. Instead, we attribute the observations to residual strain. With the help of \textit{ab initio} band structure calculations we derive a quantitative estimate for the uniaxial strain that would cause the observed valley splitting. While there are image processing methods for detecting variations of inhomogeneous strain within the field-of-view of an STM measurement \cite{Lawler2010, Walkup2018}, they are insensitive to homogeneous or nearly-homogeneous strain. This includes `accidental' residual strain that may be expected in any measurement prepared using crystal cleavage, which is quantitatively measured here fore the first time.

This work shows that even a small strain on the order of $\sim \SI{0.1}{\%}$ can lead to a dramatic rotational symmetry breaking in an energetically narrow electronic state, and in principle that this strongly symmetry breaking state can be manipulated by even weak perturbations.

\section{Results}

\subsection{STM observations of the ZrSiS surface band and Landau quantization}

Figure 1 gives an overview of the ZrSiS surface band and some of its manifestations as seen in STM measurements. The crystal structure of ZrSiS is shown in Fig. 1(a), with the cleavage plane depicted in gray. The structure belongs to the space group \textit{P}4/\textit{nmm} and cleavage results in (001) surfaces with $C_{4v}$ symmetry. Usually these surfaces are terminated by S atoms, but S contributes very little to the density of states near $E_{\mathrm{F}}$, so atomic corrugations in STM images typically correspond to the uppermost Zr layer \cite{Butler2017}. (A Si-terminated surface has also been reported \cite{Su2018}, but was not observed in this work.)

Figure 1(b) shows the band structure calculated for the S-termination using the density functional theory (DFT) framework, with the surface band shown in purple. Proceeding from the $\overline{\mathrm{X}}$ point along the $\overline{\mathrm{XM}}$ line at the Brillouin zone boundary the floating surface band starts from around $\SI{-300}{meV}$ and rises linearly to cross $E_{\mathrm{F}}$. Here it is isolated by a large energy interval from any bulk counterpart. In angle-resolved photoemission measurements it can be seen as a `v' shape along a $\overline{\mathrm{MXM}}$ cut, and in Fermi surface imaging it is shown to cross $E_{\mathrm{F}}$ on a small elliptical loop enclosing the $\overline{\mathrm{X}}$ point, giving the impression its shape is similar to a cone \cite{Fu2019}. Its behavior along the $\overline{\mathrm{\Gamma X}}$ line is somewhat more complicated however, having one branch that starts off flat from $\overline{\mathrm{X}}$ and rises slightly before turning downward to merge with the bulk nodal-loop. The partially flat bottom of the band and the downward turn are each thought to contribute a van Hove-like peak in the density of states \cite{Sankar2017,Lodge2024}. These are marked as vH1 and vH2 in Fig. 1(b). Like the lower branch, the $\overline{\Gamma}$-facing side of the cone also mixes with bulk states.

Figure 1(c) shows high energy resolution $\frac{\mathrm{d}I}{\mathrm{d}V}(V)$ curves acquired at $T = \SI{1.5}{K}$ at a cleaved ZrSiS surface. The features corresponding to vH1 and vH2 are marked, and the energy of vH1, about $\SI{330}{mV}$ below $E_{\mathrm{F}}$, can be interpreted as the energy of the band bottom. A comparison of the curves acquired under a magnetic field of $B = \SI{12}{T}$ perpendicular to the surface (solid curve), and at zero field (dashed), shows the emergence of LLs near $E_{\mathrm{F}}$. These LLs will be shown below to occur within the surface band.

Spectroscopic-imaging STM measurements can be used to image quasiparticle interference patterns resulting from scattering at defects, from which we can infer limited information about momentum-space structures. For this we measure the tunneling conductance $\frac{\mathrm{d}I}{\mathrm{d}V}(\mathbf{r},V)$ over a $150 \times \SI{150}{nm^{2}}$ field of view, first under zero magnetic field. To mitigate artifacts caused by tip-height variations we then compute the normalized conductance, $L(\mathbf{r}, V) = [\frac{\mathrm{d}I}{\mathrm{d}V}(\mathbf{r},V)]/[I(\mathbf{r}, V)/V]$, a quantity that is approximately proportional to the local density of states \cite{Kohsaka2007}. Figure 1(d) shows a representative image extracted at $V = \SI{10}{mV}$. The prominent modulations are standing waves resulting from quasiparticle interference.

Figure 1(e) displays the Fourier transform $\mathcal{F} \left[ L(\mathbf{r}, V = \SI{10}{mV} \right] $, which includes information about the momentum transfer vectors $\mathbf{q}$ that are allowed upon scattering of quasiparticles. Because the atomic lattice is also resolved, reciprocal lattice peaks appear as sharp spots, one of which is marked with a black square. The scattering signals that appear as sets of `brackets' around each reciprocal lattice peak are the result of scattering of quasiparticles from one surface band pocket to another on the opposite side of the Brillouin zone, as illustrated in Fig. 1(f) \cite{Butler2017,Su2018,Yen2021,Lodge2024}. 
Scattering within each of the surface band pockets manifests as similar sets of brackets around $q = 0$, but here the scattering signals originate from pockets in both orientations, and overlap to give the appearance of a square. (To our knowledge, scattering between one surface band pocket and one of its $\SI{90}{^{\circ}}$ rotated partners has not been reported.) All other scattering signals appearing in Fig. 1(e) are the result of scattering within the bulk nodal-loop, or between the nodal-loop and the surface bands. Some of these features have been understood in previous analyses \cite{Lodge2017,Butler2017,Su2018,Yen2021,He2021,Lodge2024} and will not be considered further here. There also exist features that have not been described previously, which appear only at low energies, and which will be the subject of a future report. For the following discussion we only note that the `bracket' signals serve as the hallmark of the surface band in quasiparticle interference observations, and the signals from sets of valleys related by a $\SI{90}{^{\circ}}$ rotation are easily distinguishable.

\begin{figure}
\centering
\includegraphics[scale=1]{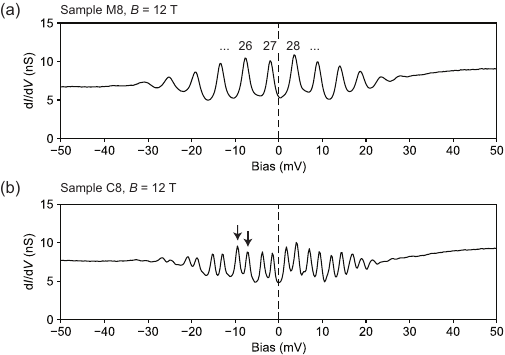}
\caption{\label{fig:2}
Observation of Landau levels. (a) A $\frac{\mathrm{d}I}{\mathrm{d}V}(V)$ curve acquired at the ZrSiS surface (sample M8, $V = \SI{50}{mV}$, $I_{\mathrm{set}} = \SI{500}{pA}$, $V_{\mathrm{mod.}} = \SI{0.25}{V}$). The index $n$ for a few of the LLs is indicated. (b) A similar $\frac{\mathrm{d}I}{\mathrm{d}V}(V)$ curve acquired on another sample (C8) showing an apparent doubling of the number of LL peaks. Two peaks that appear to be sublevels originating from the same level in (a) are marked with black arrows.
}
\end{figure}

\begin{figure*}
\centering
\includegraphics[scale=1]{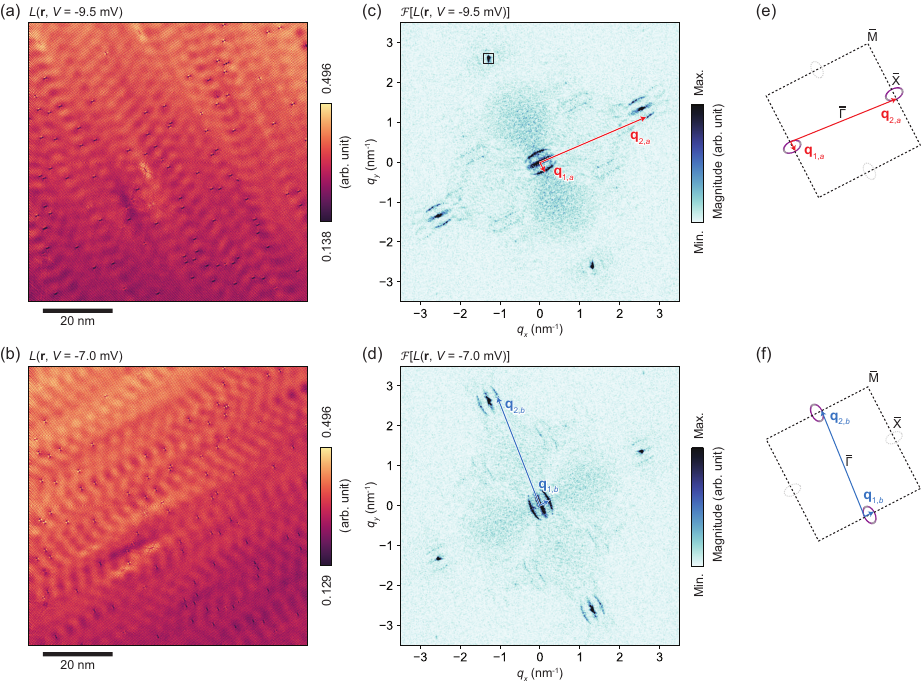}
\caption{\label{fig:3}
Imaging quasiparticle interference within valley-polarized LLs. (a) and (b) Normalized conductance images measured in the same $80 \times \SI{80}{nm^{2}}$ field-of-view, under $B = \SI{12}{T}$ ($V = \SI{50}{mV}$, $I_{\mathrm{set}} = \SI{500}{pA}$, $V_{\mathrm{mod.}} = \SI{0.3}{mV}$). The energies correspond to those of the pair of LL peaks marked with black arrows in Fig. 2(b). (c) and (d) The corresponding Fourier transform images. Prominent `bracket' scattering signals enclose only one set of reciprocal lattice peaks at each energy, and the patterns of scattering appear to be related by a $\SI{90}{^{\circ}}$ rotation. The box-like signal around $q = 0$ is also broken into the bracket-like subsets indicating intra-pocket scattering. (e) and (f) The origins of $\mathbf{q}_{1,a(b)}$ and $\mathbf{q}_{2,a(b)}$ in scattering within and between surface band pockets.
}
\end{figure*}

Focusing again on the LLs, Fig. 2(a) shows a measurement of $\frac{\mathrm{d}I}{\mathrm{d}V}(V)$ averaged over a $4 \times \SI{4}{nm^{2}}$ field of view (sample M8 - sample naming is explained in Supplementary Note 1). We first note that LLs appear only in an energy interval very near to $E_{\mathrm{F}}$. This is very similar to the appearance of LLs previously reported in the surface band of the sister compound HfSiS \cite{Jiao2018}. Some attenuation of LL peak intensity away from $E_{\mathrm{F}}$ is a nearly universal feature of LL spectroscopy measurements performed using STM \cite{Hanaguri2010, Cheng2010, Fu2014, Fu2016, Feldman2016, Jiao2018}. (One of the possible explanations is the effect of electron-phonon interactions, and in the present case LL peaks abruptly disappear upon reaching the energy of a particular phonon mode, namely a previously described $E_{g}$ mode with an energy of about $\SI{18}{meV}$ \cite{Zhou2017}.) Due to this effect, LLs may be formed in a band that spans a large energy range while only being measurable in a very narrow range. Generally the lowest LL may not then be observed and the indices of the observable LLs are not immediately known [we label them in Fig. 2(a) with the benefit of advance knowledge of the following results]. We discuss this problem and introduce a generally applicable and robust solution to it in Section D.

Figure 2(b) shows another $\frac{\mathrm{d}I}{\mathrm{d}V}(V)$ curve that was acquired under the same experimental conditions as the one shown in (a), but on a different sample (C8). This curve exhibits an apparent doubling of the number of LL peaks as compared to the measurement in (a). A reasonable guess is that there is a splitting of each LL feature into a pair. One such pair is marked in Fig. 2(b) by black arrows.

\subsection{Imaging valley-polarized LLs}

\begin{figure*}
\centering
\includegraphics[scale=1]{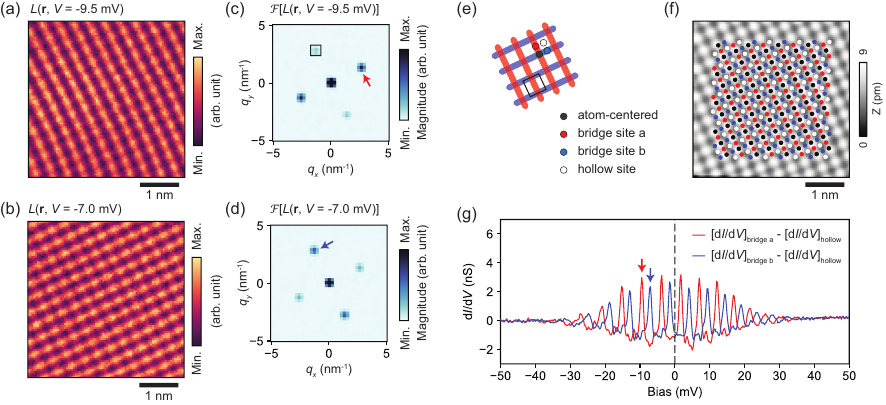}
\caption{\label{fig:4}
Obtaining valley-specific LL spectra using intra-unit-cell sampling of density of states. (a) and (b) Normalized conductance images acquired in a $4 \times \SI{4}{nm^{2}}$ field-of-view. The energies are those of the two LLs resolved in Fig. 3, and marked with black arrows in Fig. 2(b). A subtle difference can be seen in the intensity of lattice-commensurate stripes, with a $\SI{90}{^{\circ}}$ relative rotation between the two images. (c) and (d) The corresponding Fourier transform images. The difference manifests as the differing intensity around the reciprocal lattice points [one of which is marked with a black square in (c)]. The small $r$-space field-of-view results in a low $q$-space resolution, meaning that the Fourier transforms in (c) and (d) are effectively course-grained versions of those in Fig. 3(c) and 3(d) above. The quasiparticle scattering signals are now indistinguishable from the reciprocal lattice peaks, and their inverse Fourier transforms generate lattice-commensurate stripes. (e) An illustration of the resulting striped patterns, with one unit cell of the Zr square net marked in black. The sublattice sampling sites are marked with dots. (f) The sublattice sampling grids generated for this field of view (see Supplementary Note 5), shown on top of the simultaneously acquired topography image. (g) The valley-specific contributions to the conductance are obtained by subtracting the hollow site background from the bridge site curves.
}
\end{figure*}

In Fig. 3(a) and (b) we show images extracted from a $L$($\mathbf{r}$, $V$) measurement at the energy of each of the peaks in the pair marked in Fig. 2(b), specifically at $V = \SI{-9.0}{mV}$ and $V = \SI{-7.5}{mV}$, respectively. In each case the field-of-view is covered by highly coherent uniaxial modulations of the local density of states, that change orientation by $\SI{90}{^{\circ}}$ from one LL feature to the other. This difference is clarified in the corresponding Fourier transform images $\mathcal{F} \left[ L(\mathbf{r}, V) \right]$ shown in Figs. 3(c) and (d). Aside from the reciprocal lattice peaks, the most prominent features are the `bracket' signals characteristic of scattering between surface band pockets, and the signals due to scattering in the bulk bands are almost completely absent. First and foremost, this confirms that the observed LLs form in the surface band. At each energy, the inter-pocket scattering between only one pair of surface band pockets ($\mathbf{q}_{2}$) appears, and the intra-pocket scattering ($\mathbf{q}_{1}$) only within the pockets of that pair appears near $q = 0$. This shows that the splitting apparent in the $\frac{\mathrm{d}I}{\mathrm{d}V}(V)$ curve shown in Fig. 2(b) is associated with the valley degree of freedom. We now make a distinction between scattering between each of the two sets of pockets by labeling them with scattering vectors $\mathbf{q}_{1,a}$ and $\mathbf{q}_{2,a}$ within one set, and $\mathbf{q}_{1,b}$ and $\mathbf{q}_{2,b}$ within the other (in relation to the $\mathbf{a}_{0}$ and $\mathbf{b}_{0}$ lattice vectors). The origins of the scattering patterns shown in (c) and (d) are illustrated in (e) and (f), respectively. Surprisingly, scattering within the bulk bands appears to be absent at the low energies ($\left| eV \right| < \SI{10}{meV}$) where LLs are imaged, so that the $L$ and $\mathcal{F} \left[ L \right]$ images extracted between LLs are almost featureless (see Supplementary Note 2).

\subsection{Valley-selective LL spectroscopy}

We now describe a method of measuring valley-selective density of states spectra. This technique exploits the fact that in multi-valley systems, scattering vectors between valleys on opposing sides of the Brillouin zone are close to one of the reciprocal lattice vectors, and therefore their corresponding $r$-space modulations are effectively commensurate with the lattice. Due to the differing orientation of modulations derived from different valleys, a careful intra-unit-cell sampling of the measured local density of states can be used to extract the valley-specific contributions, and this can be done by measuring in only a very small field of view. (Supplementary Note 3 describes the rationale for this with numerical demonstrations for both tetragonal and triangular lattice systems.)

\begin{figure*}
\centering
\includegraphics[scale=1]{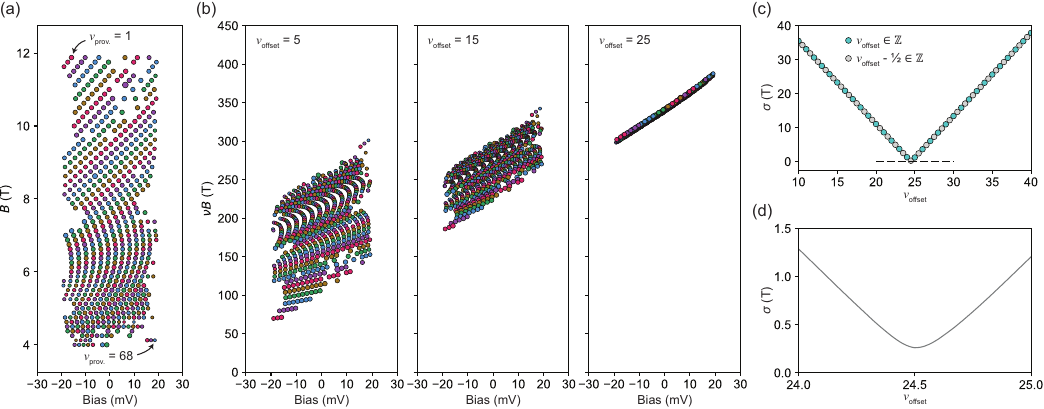}
\caption{\label{fig:5}
Obtaining LL indices and Berry phase. (a) Energies of LLs obtained (see Supplementary Note 5) at a range of different magnetic fields, from the $\frac{\mathrm{d}I}{\mathrm{d}V}(V)$ curves sampled at the $a$ sublattice sites [e.g. the red curve in Fig. 4(g)]. Points belonging to the same LL are plotted with the same color. (b) The same points shown in a plot of $\nu B$ versus energy, and the resulting plot given three trial values for $\nu_{\mathrm{offset}}$. The arrangement of points narrows as $\nu_{\mathrm{offset}}$ approaches $\sim$25. (c) The spread of points, $\sigma$, plotted against $\nu_{\mathrm{offset}}$. Gray and green markers denote values of $\nu_{\mathrm{offset}}$ that correspond to topologically trivial and non-trivial cases, respectively. $\sigma$ is minimized when $\nu_{\mathrm{offset}} = 24.5$ (topologically trivial). (d) Under the pretense that $\nu_{\mathrm{offset}}$ is not quantized but is instead a continuous variable, a precise value for $\nu_{\mathrm{offset}}$ can be obtained.
}
\end{figure*}

Figure 4 shows the implementation of this method for a $\frac{\mathrm{d}I}{\mathrm{d}V}(\mathbf{r},V)$ measurement in a $4 \times \SI{4}{nm^{2}}$ field of view. Panels (a) and (b) show images of $L$($\mathbf{r}$, $V = \SI{-9.5}{mV}$) and $L$($\mathbf{r}$, $V = \SI{-7.0}{mV}$), corresponding to the energies of the LLs visualized in Fig. 3. A very subtle $C_{2}$ symmetric and commensurate stripe pattern can be seen superimposed on the modulations of the square Zr lattice. The $\mathcal{F} \left[ L(\mathbf{r}, V) \right]$ images are shown in Figs. 4(c) and (d). The intensity of the signal around the reciprocal lattice peaks shows the same reduction to $C_{2}$ symmetry as seen in Figs. 3(c) and (d). (We discuss the correspondence between symmetry-breaking modulations seen in Figs. 3 and 4 in more detail in Supplementary Note 4.) In panel (e) we illustrate the two superimposed lattice-commensurate stripe patterns, which can be thought of as the inverse Fourier transforms of the peaks marked in panels (c) and (d). At the $r$-space points that we label as the `bridge $a$' and `bridge $b$' sites, in principle, the contribution to the local density of states from either the `$a$' or `$b$' valley dominates the signal. Figure 4(f) shows four sets of sampling sublattices, including the $a$ and $b$ bridge sites, the atom-centered (Zr-centered) sites, and the hollow sites. We average the curves over each of the four sets of sampling points in order to mitigate noise, and take the curve from the hollow sites as a baseline that we subtract from the $a$ and $b$ site curves. The resulting valley-specific curves are shown in (g), and these can be compared against the spatially averaged curve shown above in Fig. 2(b).

We perform a similar procedure to obtain valley-specific LL spectra under external magnetic fields between $B = \SI{4}{T}$ and $\SI{12}{T}$ in increments of $\SI{125}{mT}$. Using a further peak fitting procedure we can obtain values for the LL energies at each field value, building up comprehensive LL fan diagrams from which further band properties can be inferred. (See Supplementary Note 5 for all curves and for details of the sampling and fitting procedures.) The LL energies obtained in this way for the $a$ site $\frac{\mathrm{d}I}{\mathrm{d}V}(\mathbf{r},V)$ curves are shown in Fig. 5(a).

\subsection{Model-free determination of LL indices and Berry phase}

According to the Lifshitz-Onsager quantization condition, the cross-sectional area $S$ encompassed by the $n^{\mathrm{th}}$ LL orbit in $k$-space is expressed as 
\begin{equation}
S_{n} = \frac{2 \pi e B}{\hbar} \left( n + \frac{1}{2} - \frac{\gamma_{\mathrm{B}}}{2 \pi} \right),
\end{equation}
where $\hbar$ is the reduced Planck constant, $e$ is the electron charge, $n$ is an integer ($n \in \mathbb{Z}$), and $\gamma_{\mathrm{B}}$ is the Berry phase. For the procedure that follows, it is useful to define a number
\begin{equation}
\nu = n + \frac{1}{2} - \frac{\gamma_{\mathrm{B}}}{2 \pi}
\end{equation}
which we will also refer to loosely as the LL `index'. If the LL orbit encloses a topologically non-trivial point in the band structure, with $\gamma_{\mathrm{B}} = \pi$, then $\nu \in \mathbb{Z}$. If not, then $\nu - \frac{1}{2} \in \mathbb{Z}$. (The surface bands of ZrSiS are thought to emerge from a topologically trivial point.)

Generally the area $S$ is some function of energy, and Eqns. 1 and 2 establish that $S_{\nu} \propto \nu B$. Therefore, for any given dispersion relation there is some function that relates the energy of the $\nu^{\mathrm{th}}$ LL to $\nu B$:
\begin{equation}
E_{\nu} =  f(\nu B).
\end{equation}
In LL spectroscopy data such as those shown in Fig. 5(a), each LL has a distinct $E$($B$) curve that can be tracked through field-dependent measurements. If the index $\nu$ can be assigned to each observed LL, these data can be plotted in $E$-versus-$\nu B$ space, in which case (if the assignments are correct) the data points collapse onto a single curve that reflects the function in Eqn. 3. This requirement is enough to algorithmically assign a value $\nu_{i}$ to each $i^{\mathrm{th}}$ data point. We search for the unique set of values that cause all points to collapse into a curve, thereby determining the LL indices $n$ and also the Berry phase, \textit{via} Eqn. 2. (The relationship of the function in Eqn. 3 to the band dispersion will be discussed in Section E below.)

The first step used in this work is to assign a set of provisional indices $\nu_{\mathrm{prov.}}$ that will generally be wrong, but in such a way that any difference in index between any two data points is correct. This means that the error in assignment of indices is uniform over the whole set of data points, and also that the search space will only be one-dimensional. As shown in Fig. 5(a) we begin at the top-left-hand corner of the $B$-versus-bias plot, starting with $\nu_{\mathrm{prov.}} = 1$. We proceed downwards, labeling LLs until reaching the highest observed LL at $\nu_{\mathrm{prov.}} = 68$ for the 614$^{\mathrm{th}}$ data point. In this case the error, which we call $\nu_{\mathrm{offset}}$, corresponds to the number of unobservable LLs at energies lower than $\sim \SI{20}{meV}$. For the $i^{\mathrm{th}}$ point the true index is given by $\nu_{i} = \nu_{\mathrm{prov.,i}} + \nu_{\mathrm{offset}}$. Now plotting the points in $\nu B$-versus-bias space as in Fig. 5(b), we vary $\nu_{\mathrm{offset}}$ and observe which value causes the plot to collapse into a narrow curve. To aid this process we define a value $\sigma$ that characterizes the spread of the data points, and which we aim to minimize. This is obtained by sorting the data points by energy, summing the differences along the $\nu B$ axis between neighboring points, and finally dividing by the total number of data points $N$. This is written as
\begin{equation}
\sigma({\nu_{\mathrm{offset}}}) = \frac{1}{N} \sum_{i=1}^{N-1} \left| \nu_{i+1} B_{i+1} - \nu_{i} B_{i} \right|.
\end{equation}

The correct LL indices $\nu_{i}$ are found where the choice of $\nu_{\mathrm{offset}}$ minimizes $\sigma$, i.e.:
\begin{equation}
\nu_{i} = \nu_{\mathrm{prov.}, i} + \operatorname*{argmin}_{\nu_{\mathrm{offset}}} \sigma
\end{equation}

The number $\nu_{\mathrm{offset}}$ also encodes the unknown Berry phase that contributes to $\nu$ (see Eq. 2) and this can take a value of zero or $\pi$ so that either $\nu_{\mathrm{offset}}$ or $\nu_{\mathrm{offset}} - \frac{1}{2}$ is an integer. Figure 5(c) shows the variation of $\sigma$ when $\nu_{\mathrm{offset}}$ is incremented in both cases, and that it is minimized when $\nu_{\mathrm{offset}} = 24.5$. This means that any given LL index belongs to $\mathbb{Z} + \frac{1}{2}$, corresponding to the case of zero Berry phase. This is consistent with previous observations of quantum oscillations in the ZrSiS surface band \cite{Liu2021}. We can confirm the above result by adopting the pretense that $\nu_{\mathrm{offset}}$ is not quantized but is instead a continuous variable as shown in Fig. 5(d). $\sigma$ is minimized almost exactly at $\nu_{\mathrm{offset}} = 24.5$. The lowest observed LL index in the data set shown in Fig. 5(a) is $n = 25$, and the highest is $n = 92$.

\subsection{Estimation of valley-dependent energy shift}

To better understand and quantify the valley-splitting seen in Figs. 3 and 4, we anticipate plotting the LL energy data in Fig. 5(a) and (b) as a dispersion curve $E$($k$). Constructing a model for the dispersion requires that we make an assumption about the shape of the $k$-space region covered by $S_{\nu} \propto \nu B$ (i.e. the shape of the band's iso-energetic contour), which gives the form of $k$($\nu B$). If we assume an isotropic dispersion, a circular LL orbit with area $S_{\nu}$ has a radius 
\begin{equation}
k_{\nu} = \sqrt{ \frac{ 2 e \nu B }{ \hbar }}.
\end{equation}

Having determined $\nu$ for each observed LL, we take the data shown in Fig. 5(b) for valley \textit{a}, and obtain an $E$($k$) plot, which is shown in Fig. 6(a). Going through a parallel process to obtain LL energies for valley \textit{b} (see Supplementary Note 5), we plot the corresponding curve in Fig. 6(b). This yields a pair of curves that are each approximately linear, and this is in good agreement with the expected dispersion of the surface band near $E_{\mathrm{F}}$, which is seen in the DFT calculations shown in Fig. 1(b) as well as in previous reports \cite{Topp2017,Fu2019}.

\begin{figure}[b]
\centering
\includegraphics[scale=1]{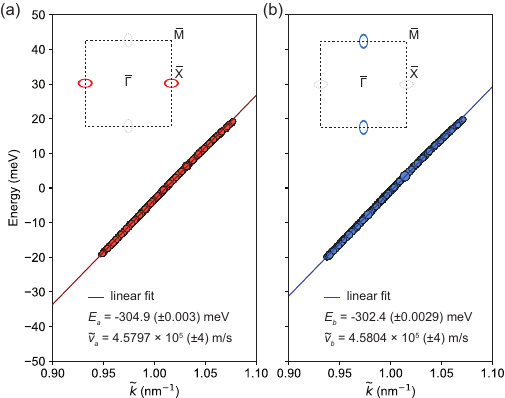}
\caption{\label{fig:6}
Valley-selective dispersion of the floating surface band near $E_{\mathrm{F}}$. (a) A dispersion curve obtained for the $a$ valley, using LL indices with $\nu_\mathrm{offset} = 24.5$ (from Fig. 5). The solid red line is a linear fit to all data points, with the resulting $E_{a}$ and average velocity $\tilde{v}_{a}$ also shown. (b) The corresponding plot and linear fit result for the points extracted from the $b$ sublattice sites and following the same procedure as shown in Fig. 5, which yields $\nu_\mathrm{offset} = 23.5$. A comparison of the linear fitting results shows a difference $\left| E_{b} - E_{a} \right| = \SI{2.5}{meV}$, and a negligible difference in Fermi velocities.
}
\end{figure}

We perform linear fits using $E(k) = E_{a(b)} + \tilde{v}_{a(b)} \tilde{k}$ for valleys $a$ and $b$ respectively. Because the iso-energetic contours of the real surface bands are not circular but elliptical, we have specified $\tilde{v}$ as the absolute band velocity on a circular orbit with the same area as the real band contour. Likewise, $\tilde{k}$ is the radius of the equivalent circular orbit. The linear fitting results are shown in Figs. 6(a) and (b). We find that $E_{a}$ and $E_{b}$ differ by $\SI{2.5}{meV}$ [c.f. Figs. 3(a) and (b)], and the difference between $\tilde{v}_{a}$ and $\tilde{v}_{b}$ is negligible.

For later reference it is useful to construct a model that allows us to simulate the expected LL spectrum for a linear dispersion, and later to include a parameter to represent the valley splitting. We start by considering how the area $S_{\nu} \propto \nu B$ varies with energy, or \textit{vice versa}, which depends on the form of $E$($k$). Generally $E_{\mathrm{\nu}}$ scales as $\left( \nu B \right) ^{\alpha}$, with $\alpha = 1$ for a parabolic band, $\alpha = \frac{1}{2}$ for a linear band \cite{Li2009}, and other values for other, more exotic cases \cite{Dietl2008}. Adopting $\alpha = \frac{1}{2}$ for reasons given above, and using Eqn. 6, we have
\begin{equation}
E_{\nu} = E_{0} + \tilde{v} \sqrt{2e \hbar \nu B}.
\end{equation}
where $E_{0}$ is the energy of the bottom of the conical band. We do not propose or require that this model represents the overall dispersion of the ZrSiS surface band. It does provide a very effective model to understand the LL dispersion in the narrow range of $\pm \SI{20}{meV}$ around $E_{\mathrm{F}}$, as will be shown below.

Fig. 7(a) displays the surface bands modeled as a set of cones of elliptic cross-section located around the $\overline{\textrm{X}}$ points of the surface Brillouin zone. Although it only partially resembles the structure shown in Fig. 1(b), it will be shown below to very accurately capture the key band properties near $E_{\mathrm{F}}$, which are relevant to the observed LL formation.

In Fig. 7(b) we illustrate the Landau quantization of the surface band cones according to Eq. 7. Figure 7(c) shows the simulated LL spectrum, as well as a zoom-in view of the energy interval of observational interest. For these plots we adopt realistic values for $E_{0}$ and the band velocity \cite{Fu2019}, and assume the topologically trivial case, and the result very closely resembles the measured spectrum in Fig. 2(a).

A valley degree-of-freedom can be added to the model described by Eqn. 7. The ways of doing this could have been i) to add a rigid, valley dependent energy shift $\delta_{\mathrm{valley}}$, or ii) to include a valley dependent modification of the band velocity (with a third option being a combination of these effects). The results of the linear fitting above show that a valley dependent difference in velocities can be safely excluded, and the total LL spectrum is then given by
\begin{equation}
E_{\nu, \pm} = E_{0} + \tilde{v} \sqrt{2e \hbar \nu B} \pm \delta_{\mathrm{valley}}.
\end{equation}

Referring back to Fig. 6, we can recognize that $E_{a} = E_{\nu=\frac{1}{2},-}$ and $E_{b} = E_{\nu=\frac{1}{2},+}$. Valley-dependent LL spectra can be simulated for comparison with observation, and we show this in Fig. 8 below.

\begin{figure}
\centering
\includegraphics[scale=1]{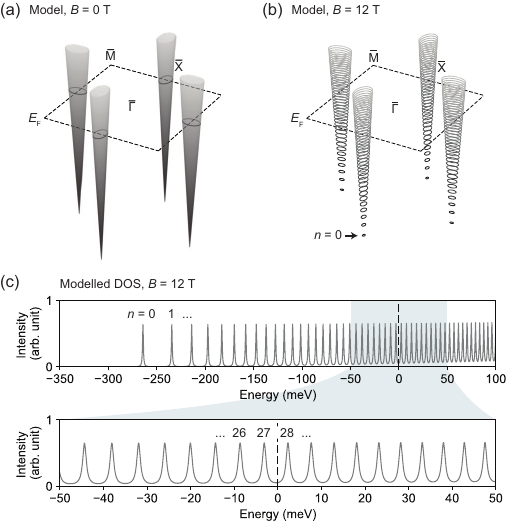}
\caption{\label{fig:7}
Simplified model for understanding LL spectra in the surface band. (a) A model comprising a cone-shaped band at each $\overline{\textrm{X}}$ point of the surface Brillouin zone. (b) A depiction of Landau quantization under a field of $B = \SI{12}{T}$. (c) Modeled density of states spectrum at $B = \SI{12}{T}$ and a zoom-in view of the narrow interval around $E_{\mathrm{F}}$ where LLs can be resolved in measurements and where the model has some validity [see Fig. 2(a)]. Each LL is represented as a Lorentzian lineshape.
}
\end{figure}

\begin{figure*}
\centering
\includegraphics[scale=1]{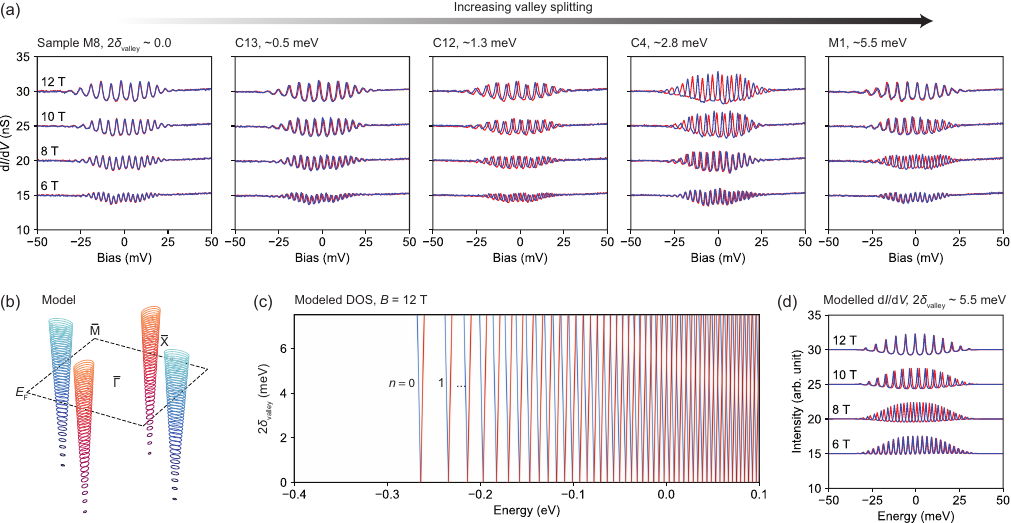}
\caption{\label{fig:8}
Sample dependence of valley splitting. (a) Valley-selective $\frac{\mathrm{d}I}{\mathrm{d}V}(V)$ curves at various magnetic fields for five different samples, sorted in order of the valley splitting energy, from lowest to highest. The asymmetry seen at $B = \SI{12}{T}$ is greatest for sample C4. For the sample with the highest splitting, where $B \approx \SI{12}{T}$ the valley splitting is similar to the energy spacing between LLs. At $B = \SI{12}{T}$, LLs of different filling, from different valleys, coexist at the same energy. (b) The cone model for the floating surface band. Here the two inequivalent sets of valleys are depicted in blue and red. (c) Density of states modeled using Eqn. 8 for the two LL series under $B = \SI{12}{T}$, plotted with varying valley splitting $2\delta_{\mathrm{valley}}$. (d) Simulated $\frac{\mathrm{d}I}{\mathrm{d}V}(V)$ curves at the same magnetic fields as for (a), for a splitting of $2\delta_{\mathrm{valley}} = \SI{5.5}{mV}$. (An envelope function is applied for ease of comparison.)
}
\end{figure*}

\subsection{Sample dependence of valley splitting}

As suggested by the difference in the $\frac{\mathrm{d}I}{\mathrm{d}V}(V)$ curves shown in Figs. 2(a) and 2(b), there is significant sample dependence of the valley splitting. Figure 8(a) shows series of valley-selective $\frac{\mathrm{d}I}{\mathrm{d}V}(V)$ curves acquired from five selected samples. From left to right, the splitting is seen to range from zero to $\sim \SI{5.5}{meV}$.

Assuming that, in Eq. 7, $E_{0}$ and $\tilde{v}$ have negligible sample dependence, it seems possible to simply read out the value of $\delta_{\mathrm{valley}}$ for LL spectra observed in any sample without the process detailed in Figs. 5 and 6 above. The top row of curves in Fig. 8(a) show that this is not the case, however. If $\delta_{\mathrm{valley}}$ is large enough, at some point it matches the LL spacing and the two curves become indistinguishable, resembling the case of $\delta_{\mathrm{valley}}$ = 0. Pairs of curves collected at more than one magnetic field are needed to unambiguously determine $\delta_{\mathrm{valley}}$. The values given in Fig. 8(a) are obtained by measuring a short series of curves that capture enough field dependent information, and using a trial-and-error process of simulating the corresponding field dependence for a given value of $\delta_{\mathrm{valley}}$.

Figure 8(b) illustrates the model underlying the LL spectra described by Eq. 7, and 8(c) shows the resulting valley dependent spectra simulated for varying $\delta_{\mathrm{valley}}$ under $B = \SI{12}{T}$. Each observation at $\SI{12}{T}$ shown in (a) corresponds to a horizontal cut through the simulation in (c). Given this, a trial-and-error search along the $2\delta_{\mathrm{valley}}$-axis of the simulation can be used to infer the splitting in the measured spectra, and an example of this is given in Fig. 8(d). The simulated spectra at chosen magnetic fields using $2\delta_{\mathrm{valley}} = \SI{5.5}{meV}$ yields a very good match to the spectra shown in Fig. 8(a) for sample M1. The success of this procedure attests to the validity of the simple model illustrated in Fig. 8(b), that yields the LL spectrum described by Eq. 7.

\section{Summary and Discussion}

We have used STM to observe LLs at the surface of ZrSiS under high magnetic fields. We have seen that the energies of the observed LLs closely match those of the predicted LL spectrum for the floating surface band, if it is modeled simply as a set of cones, each located at one of the $\overline{\mathrm{X}}$ points. But we also find that most samples exhibit an unexpected splitting of the LL spectrum. Quasiparticle interference imaging shows that the splitting manifests within the valley degree of freedom, indicating that each pair of valleys supports an independent LL spectrum related by a rigid energy difference of $2\delta_{\mathrm{valley}}$. The valley splitting is seen to be sample-dependent, ranging from zero to a few $\SI{}{meV}$.

Due to this effect, a measurement at any particular energy will generally probe a LL in at most one of the sets of valleys, as depicted in Fig. 3. We can switch between states of opposite valley-polarization by adjusting the energy by only a few $\SI{}{meV}$. Also, a LL of either valley polarization can be brought to $E_{\mathrm{F}}$ using a small ($\sim \SI{100}{mT}$) adjustment of the magnetic field. In principle the valley polarization is therefore amenable to detection in a suitable quantum oscillation measurement \cite{Liu2021}. A noteworthy complication is that a relatively large valley splitting can coincide with the energy interval between two LLs. This situation can be seen in Fig. 8(a), in the $\SI{12}{T}$ measurement for sample M1, where we have the unusual case of a single spectroscopic feature composed of LLs with differing indices.

\begin{figure}
\centering
\includegraphics[scale=1]{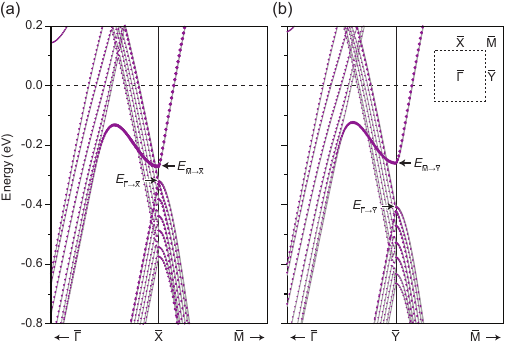}
\caption{\label{fig:9}
Calculation of strain effects on the surface bands. (a) Surface band structure around the $\overline{\mathrm{X}}$ point for the case of a $\SI{1}{\%}$ uniaxial strain (elongating the lattice vector $\mathbf{b}_{0}$), calculated using DFT. (b) Surface band structure around the $\overline{\mathrm{Y}}$ point of the strained Brillouin zone (shown in the inset). There are two features in the band structure at each point that could each be identified with the bottom of one cone in the simplified model investigated above.
}
\end{figure}

The application of a magnetic field is not thought to be responsible for the lifting of valley degeneracy and we observe no influence of magnetic field on the valley splitting. This can be verified by considering the collapse of a set of measured LL energies into a narrow curve, as shown in Fig. 5. Because this set includes points obtained under very different fields, the collapse could not occur if there was a significant field dependent contribution to the band energy. Though a lifting of the valley degeneracy under magnetic field is not expected, lifting of the two-fold spin degeneracy within each valley is expected. However, surprisingly it was not observed. The Zeeman splitting for free electrons under a field of $\SI{12}{T}$ would be $\sim \SI{1.4}{meV}$, and no such splitting appears in either the spatially averaged or valley-resolved $\frac{\mathrm{d}I}{\mathrm{d}V}(V)$ curves. Even a smaller Zeeman splitting (for Land\'{e} factor $g < 2$) might be expected to cause a field-dependent contribution to the apparent width of the LL peaks, but this also was not observed (see Supplementary Note 6). The apparent absence or diminution of the Zeeman effect is noteworthy but is beyond the scope of this work.

The sample dependence of $\delta_{\mathrm{valley}}$, and especially the observation of $\delta_{\mathrm{valley}} = 0$ [Fig. 8(a), sample M8] rules out any spontaneous symmetry breaking of the electronic states. Instead, the lifting of valley degeneracy is likely caused by residual strain. In Fig. 9 we show the calculated electronic band structure for $\SI{1}{\%}$ strained ZrSiS, focusing on the regions around the $\overline{\mathrm{X}}$ and $\overline{\mathrm{Y}}$ points of the strained Brillouin zone. At each high-symmetry point we can find two features that can be associated with the bottom of the cones in our simplified model. In the real band structure, if we start from the Fermi contour around $\overline{\mathrm{X}}$ (or $\overline{\mathrm{Y}}$), the bands extending downwards on the $\overline{\mathrm{M}}$-facing and $\overline{\mathrm{\Gamma}}$-facing sides of the manifold reach the respective high symmetry point at different energies. In Fig. 9 we label these as $E_{\overline{\mathrm{M}} \rightarrow \overline{\mathrm{X}}(\overline{\mathrm{Y}})}$ and $E_{\overline{\Gamma} \rightarrow \overline{\mathrm{X}}(\overline{\mathrm{Y}})}$, respectively. The former of these two features is $\SI{9.8}{meV}$ higher at $\overline{\mathrm{Y}}$ than at $\overline{\mathrm{X}}$, while the latter is $\SI{87.3}{meV}$ lower at $\overline{\mathrm{Y}}$ than at $\overline{\mathrm{X}}$. If we take the average of these shifts in order to draw an estimate of the corresponding valley splitting in our conical model, this gives $2 \delta_{\mathrm{valley}} \approx$ $\SI{40}{meV}$ resulting from $\SI{1}{\%}$ uniaxial strain. Experimentally, a change in valley splitting of $< \SI{1}{meV}$ can easily be resolved, meaning that the strain can be estimated to a precision of about $\SI{0.025}{\%}$. The largest valley splitting that was observed in this work corresponds to a uniaxial strain of about $\SI{0.14}{\%}$. The average valley splitting observed for seven samples was $\SI{1.97}{meV}$, corresponding to an average uniaxial strain of $\SI{0.049}{\%}$. As well as significant sample dependence, the valley splitting also appears to exhibit some spatial dependence, and we discuss this in Supplementary Note 7.

Note that only the $\left< 100 \right>$ component of strain can be addressed in this work, and this is due to the location of the surface band valleys at the $\overline{\mathrm{X}}$ points. The $\left< 110 \right>$ component of strain would only be observable in a system with valleys located at the $\overline{\mathrm{M}}$ points.



An immediate conclusion of the above observation is that residual strain on the order of $\sim \SI{0.1}{\%}$ is probably a ubiquitous feature of STM measurements on cleaved crystals even when deliberate steps are taken to reduce it (see Supplementary Note 1), and it should be considered when using STM to investigate symmetry breaking electronic phenomena. This leads us to caution that this effect could also be significant not only in Landau-quantized systems, but also in intrinsic flatband systems such as the kagom\'{e} metals, leading to apparently spontaneous (but actually strain-induced) symmetry breaking in the local density of states.

The observations of valley-selective LL spectra shown in Fig. 4 demonstrate a method of using STM that should be helpful for directly addressing the valley degree of freedom using only a very local $r$-space measurement covering only a relatively few ($\sim$100) unit-cells. Although the procedure is used for a square lattice system here, it should also be expected to work straightforwardly for a triangular lattice, and therefore may prove useful for elucidating new valley-related phenomena in a broad range of material systems.

We also introduce a procedure, summarized in Fig. 5, for determining the indices of observed LLs and the Berry phase around a LL orbit in a situation where none of these are initially known. In principle there are systems such as isolated monolayer materials where gating may help to bring the lowest LL to $E_{\mathrm{F}}$, and perhaps into the energy range where it is easiest to measure, and the established methods can be used \cite{Hanaguri2010}. This technique would avoid the need for gating in that case and can also apply to non-gate-able systems such as bulk metals. Another merit of this technique is that it is agnostic to the underlying band dispersion of the system under investigation. It would have been equally useful if the dispersion had been quadratic, for example, or indeed if it had any arbitrary dispersion. For this reason it is robust against unexpected band distortions due to many-body renormalization effects, and useful for characterizing them. (We discuss some aspects of the method further in Supplementary Note 8.)

In summary, the above results elucidate how in a multi-valley band structure, a residual strain on a scale of only $\sim \SI{0.1}{\%}$ can lead to extremely valley-polarized LLs, and provide detailed microscopic insights for both strain- and valley-engineering of quantum materials. With a narrower focus on microscopic observations of symmetry breaking states, we also suggest that the strain inferred here is typical of a generic STM measurement, and could impact how measurements of symmetry breaking in other narrow spectroscopic features are interpreted, such as those associated with flatbands or other contributors of van Hove-like signatures.

\section{Materials and Methods}

\subsection{Crystal synthesis}

ZrSiS polycrystals were synthesized by the solid-state reaction using the elemental powders (Zr: $\SI{98.8}{\%}$, Si:  $\SI{99.99}{\%}$, and S: $\SI{99.995}{\%}$) with the compositional ratio in evacuated quartz tubes at $\SI{1100}{^{\circ}C}$ for 100 hours. Subsequent crystal growth was carried out by the chemical vapor transport method using the polycrystals with $\SI{10}{wt\%}$ of iodine ($\SI{99.9995}{\%}$) in evacuated quartz tubes for 100 hours at $\SI{1100}{^{\circ}C}$ for the source zone and $\SI{1000}{^{\circ}C}$ for the growth zone. The crystal structure, crystal orientation, and chemical composition of the obtained single crystals were evaluated by X-ray diffraction, X-ray Laue back reflection, and X-ray fluorescence techniques, respectively.

\subsection{STM measurements}

Crystals were glued using conducting epoxy to either a Cu or Mo sample plate, which was then fixed to a BeCu alloy holder (see Supplementary Note 1). After loading samples into an ultra-high vacuum chamber ($P \sim \SI{e-10}{Torr}$), surfaces were prepared for STM measurements by cleaving them at about $\SI{77}{K}$, before insertion into a modified Unisoku USM1300 low-temperature STM system held at $T = \SI{1.5}{K}$ \cite{Hanaguri2006}. STM measurements were performed using electro-chemically etched tungsten tips, which were characterized and conditioned using field ion microscopy followed by repeated mild indentation at a clean Cu(111) surface. Special care was taken to form a tip apex that was both as sharp and as isotropic as possible, which enables effective intra-unit-cell sampling of $\frac{\mathrm{d}I}{\mathrm{d}V}(V)$ data as described in Fig 4. Tunneling conductance was measured using the lock-in technique with bias modulation of frequency $f_{\textrm{mod}} = \SI{617.3}{Hz}$ and an amplitude $V_{\mathrm{mod.}}$ specified in the caption describing each measurement. 

The external magnetic field was applied along the $z$-axis of the STM system, and for each sample the tilt of the surface was characterized using STM. It was confirmed that the surface-perpendicular vector was no more than $\sim \SI{1}{^{\circ}}$ away from the $z$-axis. This ensures that deviation of the magnetic field strength from the nominal value was negligible.

Conductance maps and their Fourier transforms are plotted using perceptually uniform colormaps \cite{Thyng2016}.


\subsection{DFT calculations}

The calculations of the electronic structure of the ZrSiS were performed using a slab construction, using density functional theory (DFT) as implemented in the Vienna \textit{ab initio} simulation package (VASP)~\cite{Kresse1996a,Kresse1996b}.
All the calculations were performed using the projector-augmented wave (PAW)~\cite{Blochl1994} pseudopotential with the generalized gradient approximation (GGA) in the form of Perdew-Burke-Ernzerhof (PBE)~\cite{Perdew1996,Kresse1999}.
The plane-wave cutoff energy was $\SI{500}{eV}$ and a $\Gamma$-centered $4\times 4 \times 1$ $k$-mesh was used to describe the electronic structure.
The valence orbital set was $4s^24p^65s^24d^2$ for Zr, $3s^23p^2$ for Si, and $3s^23p^4$ for S.
The 7-layer ZrSiS slab with S terminations was modeled utilizing the supercell approach, with the separations between the neighboring slabs being about $\SI{21}{\mathrm{\AA}}$.
We used the experimentally obtained lattice parameter  $a_{0} = \SI{3.544}{\mathrm{\AA}}$, as well as the experimental atomic position for the ZrSiS slab \cite{Sankar2017}. The definition of the strain applied is the discrepancy between lattice constant $a_{0}$ and $b_{0}$ (elongating $b_{0}$).

\section*{Acknowledgements}
We are grateful to  T. Machida and M. Naritsuka for assistance, and to M. Kawamura, P. J. Hsu, T.-M. Chuang and A. Rost for helpful discussions. This work was supported by JST CREST Grant No. JPMJCR16F2 and No. JPMJCR20B4, and also by a Grant-in-Aid for Scientific Research on Innovative Areas `Quantum Liquid Crystals' (KAKENHI Grant No. JP19H05824 and No. JP19H05825), and for `Science of 2.5 Dimensional Materials’ (KAKENHI Grant No. JP21H05236), for Scientific Research (A) (KAKENHI Grant No. JP21H04652), and for Challenging Research (Pioneering) (KAKENHI Grant No. JP21K18181) from JSPS of Japan. C.J.B. acknowledges support from RIKEN's Programs for Junior Scientists. M.-C. J. acknowledges support from RIKEN's IPA program.

\section*{Data availability}
The data that support the findings presented here are available from the corresponding authors upon reasonable request.

\end{document}


\title{Supplementary Material for\\Valley polarization of Landau levels in the ZrSiS surface band \\ driven by residual strain}

\author{Christopher J. Butler}
\email{christopher.butler@riken.jp}
\affiliation{RIKEN Center for Emergent Matter Science, 2-1 Hirosawa, Wako, Saitama 351-0198, Japan}

\author{Masayuki Murase}
\affiliation{Laboratory for Materials and Structures, Tokyo Institute of Technology, Kanagawa 226-8501, Japan}

\author{Shunki Sawada}
\affiliation{Laboratory for Materials and Structures, Tokyo Institute of Technology, Kanagawa 226-8501, Japan}

\author{Ming-Chun Jiang}
\affiliation{RIKEN Center for Emergent Matter Science, 2-1 Hirosawa, Wako, Saitama 351-0198, Japan}
\affiliation{Department of Physics and Center for Theoretical Physics, National Taiwan University, Taipei 10617, Taiwan}

\author{Daisuke Hashizume}
\affiliation{RIKEN Center for Emergent Matter Science, 2-1 Hirosawa, Wako, Saitama 351-0198, Japan}

\author{Guang-Yu Guo}
\affiliation{Department of Physics and Center for Theoretical Physics, National Taiwan University, Taipei 10617, Taiwan}
\affiliation{Physics Division, National Center for Theoretical Sciences, Taipei 10617, Taiwan}

\author{Ryotaro Arita}
\affiliation{RIKEN Center for Emergent Matter Science, 2-1 Hirosawa, Wako, Saitama 351-0198, Japan}
\affiliation{Department of Physics, University of Tokyo, 7-3-1 Hongo Bunkyo-ku, Tokyo 113-0033, Japan}

\author{Tetsuo Hanaguri}
\email{hanaguri@riken.jp}
\affiliation{RIKEN Center for Emergent Matter Science, 2-1 Hirosawa, Wako, Saitama 351-0198, Japan}

\author{Takao Sasagawa}
\affiliation{Laboratory for Materials and Structures, Tokyo Institute of Technology, Kanagawa 226-8501, Japan}
\affiliation{Research Center for Autonomous Systems Materialogy, Tokyo Institute of Technology, Kanagawa 226-8501, Japan}

\maketitle

\section*{Supplementary Note 1 - Sample nomenclature and efforts to reduce strain} 

Upon the first observation of an apparent reduced-symmetry electronic state of the kind shown in Fig. 3 of the main text, it needs to be determined whether or not the symmetry breaking is spontaneous. `Smoking gun' evidence for or against the spontaneity of the symmetry breaking might be obtained by observing whether it shows ferroic-like behavior under externally applied strain that is smoothly swept through zero. In principle this could be achieved using recently developed instruments \cite{Hicks2014}, but in absence of this we established the (non-)spontaneity of the symmetry breaking state simply by measuring many samples. Also, instead of deliberately introducing strain, for some samples we took steps to reduce strain caused by differential thermal contraction of the crystal and sample plate.

Short series of magnetic field dependent, valley-selective d$I$/d$V$($V$) spectra similar to those shown in Fig. 8(a) of the main text were acquired on a total of seven samples. Ordinarily our samples are fixed using epoxy (EpoTek H20E) to thin Cu plates and screwed to BeCu alloy holders using two Mo screws [see Fig. S1(a) and S1(b)]. Anticipating the role of strain we attempted to alter this setup by instead fixing three of the samples to Mo plates and reducing the number of screws from two to one [see Fig. S1(a) and S1(c)]. The choice of Mo as the sample plate results in a better match of the thermal expansion coefficient between ZrSiS and the sample plate, which might reduce strain due to cooling of samples from $\SI{300}{}$ to $\SI{1.5}{K}$ [see Fig. S1(d)]. The use of only one screw to secure the Mo sample plate is intended to reduce uniaxial strain due to differing thermal contraction between the sample plate and the BeCu holder.

\begin{figure*}
\centering
\includegraphics[scale=1]{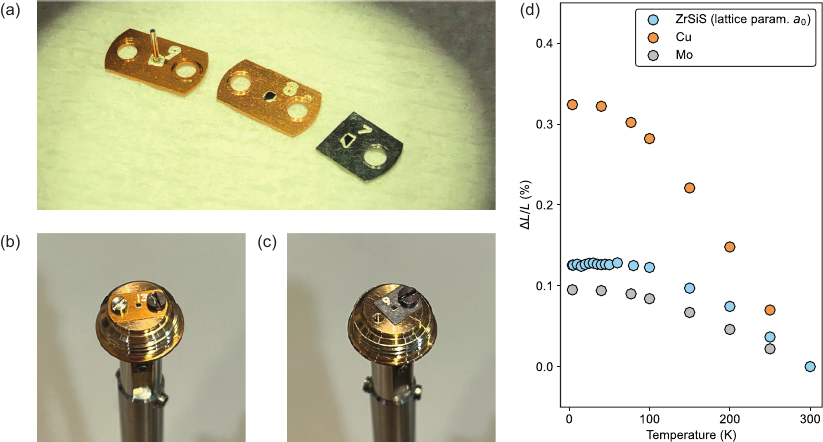}
\caption{\label{fig:1}
Efforts to reduce residual strain in cleaved ZrSiS samples. (a) Examples of ZrSiS samples before (left) and after (center and right) cleavage. The samples on the left and in the center are glued to Cu plates, and the sample on the right is glued to a Mo plate. (b) and (c) Sample plates mounted by Mo screws to BeCu alloy sample holders. (d) Measurement of thermal expansion as observed in the lattice parameter $a_{0}$ for ZrSiS with varying temperature, as well as values for Cu and Mo \cite{EkinBook}. The thermal contraction of ZrSiS is reasonably well matched to that of Mo.
}
\end{figure*}

\begin{figure*}
\centering
\includegraphics[scale=1]{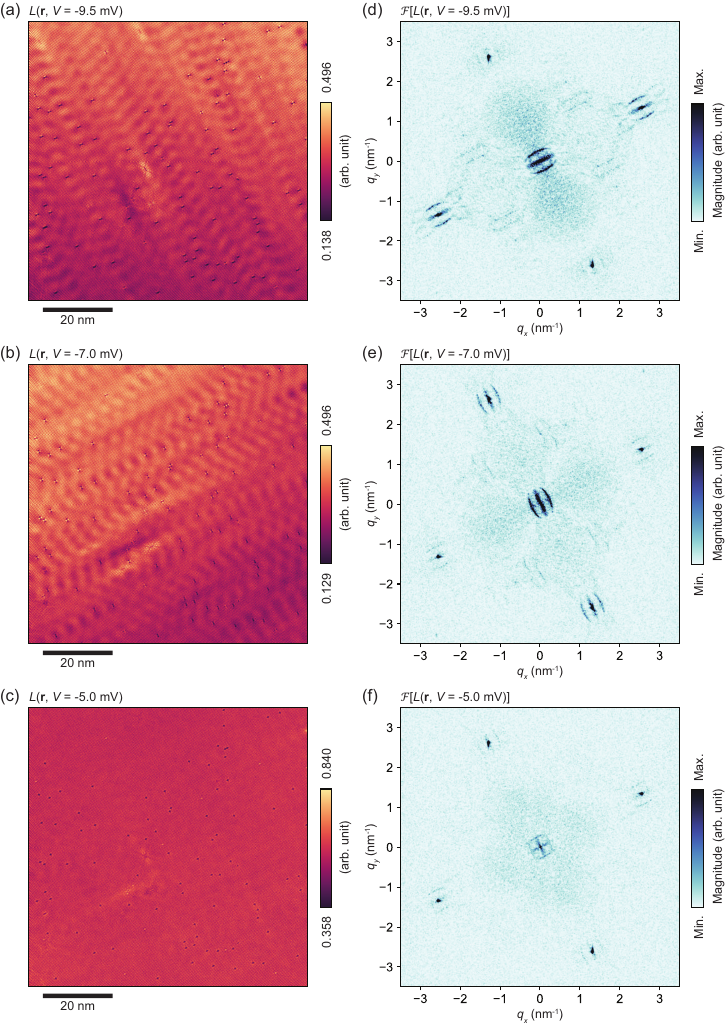}
\caption{\label{fig:2}
Conductance imaging at and between valley-polarized LLs. (a) and (b) Normalized conductance images of the $26^{\mathrm{th}}$ LL of each valley, as shown in Fig. 3 of the main text. (c) The normalized conductance image extracted at $V = -\SI{5}{mV}$, in between the $26^{\mathrm{th}}$ and $27^{\mathrm{th}}$ LLs. (d), (e) and (f) The corresponding Fourier transform images. As expected, the intensity of the `bracket' signals for surface band scattering appear significantly weaker between LLs. Surprisingly, the scattering signals associated with the bulk bands are absent for all images.}
\end{figure*}

We identify these various samples with a number that is prefixed by `C' for those mounted on Cu and `M' for those mounted on Mo. Although the origin of residual strain in STM measurements is not well understood, it might be expected that the difference between the two sample setups would lead to some systematic difference in observed valley splitting. However, no such difference was found. Indeed, the sample with the smallest observed valley splitting, and the one with the largest splitting were both mounted on Mo [see Fig. 8(a) of the main text].

\section*{Supplementary Note 2 - Imaging of quasiparticle interference at and between Landau levels}

In Fig. S2 we show normalized conductance images $L$($\mathbf{r}$) and their Fourier transforms $\mathcal{F} \left[ L(\mathbf{r}) \right]$ for the $26^{\mathrm{th}}$ LL of each valley, as shown in Fig. 3 of the main text. Here we also show the corresponding images extracted at an energy between the $26^{\mathrm{th}}$ and $27^{\mathrm{th}}$ LLs. Because the LLs are formed within the surface bands, between LLs the intensity of the `bracket' signals for surface band scattering appear significantly weaker. Surprisingly, scattering within the bulk bands appears to be absent at the low energies ($\left| eV \right| < \SI{10}{meV}$) where LLs are imaged so that, aside from the atomic corrugation and Bragg peaks, the $L$ and $\mathcal{F} \left[ L \right]$ images extracted between LLs are almost featureless. This phenomenon of absent bulk band scattering will be discussed in a future report.

\newpage

\section*{Supplementary Note 3 - Intra-unit-cell sampling for valley-selective tunneling spectroscopy in tetragonal and triangular lattice systems.}

In Figure S3 below we show with a numerical demonstration how sub-lattice sampling method can apply for both triangular lattice systems as well as for the square lattice system as demonstrated in the main text. We do this by simulating quasiparticle interference patterns [using a simple joint-density-of-states (JDOS) approximation] in both square and triangular multi-valley systems, and then taking the inverse Fourier transforms of the patterns. The key is to label each valley (which we do using different colors in Fig. S3), and keep track of each valley’s contribution in the \textit{r}-space motif resulting from the inverse Fourier transform. Then, as shown in Fig. S3(j) for the triangular lattice, we can identify the \textit{r}-space sublattice locations where careful sampling would enable valley-selective spectroscopy measurements.

\begin{figure*}
\centering
\includegraphics[scale=1]{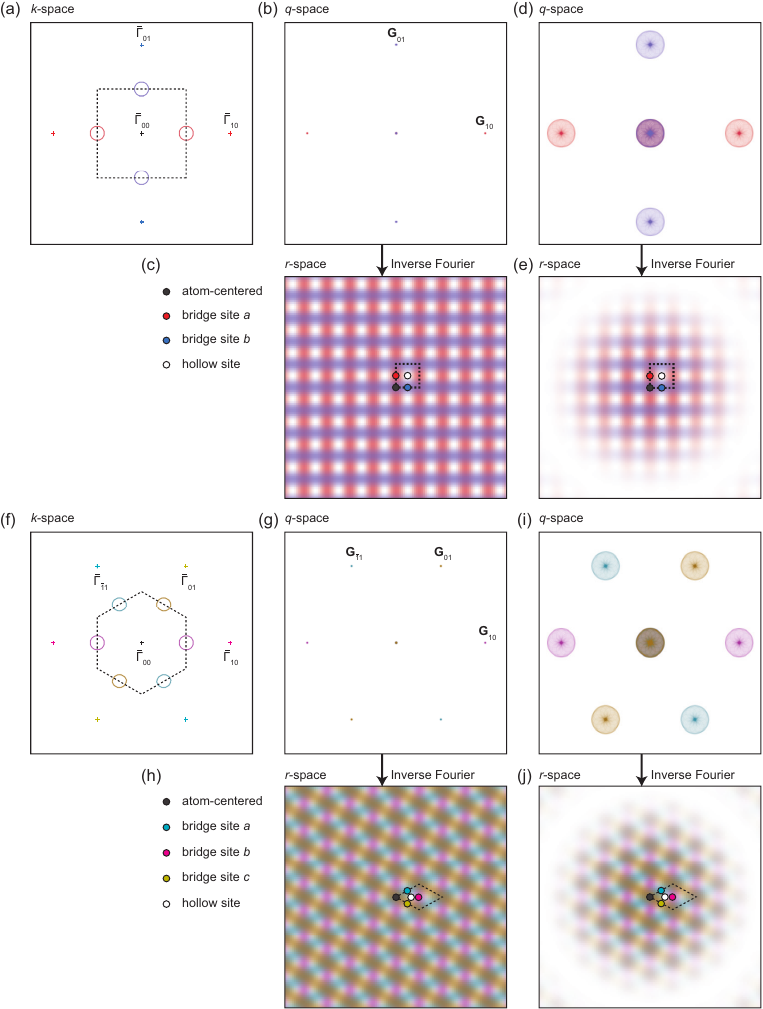}
\caption{\label{fig:3}
Sublattice sampling for valley-selective spectroscopy for square and triangular lattices. (a) An iso-energetic band contour for a multi-valley square lattice system. Red and blue signify the valley identities. (b) The Bragg peaks and (c) their inverse Fourier transforms. One unit-cell is depicted with the sampling sites enabling valley-selectivity indicated. (d) Each band contour's JDOS and (e) inverse Fourier transform. Sublattice sites for valley-selectivity are marked. (f) Band contours for a triangular lattice system, where cyan, magenta, and yellow colors signify the valley identities. (g) Bragg peaks (h) and their inverse Fourier transforms. (i) The JDOS for each contour and (j) the inverse Fourier transform. The sampling sites for valley-selectivity are marked.
}
\end{figure*}

\clearpage

\section*{Supplementary Note 4 - Correspondence between symmetry breaking modulations in Figs. 3 and 4 of the main text.}

\begin{figure*}[b]
\centering
\includegraphics[scale=1]{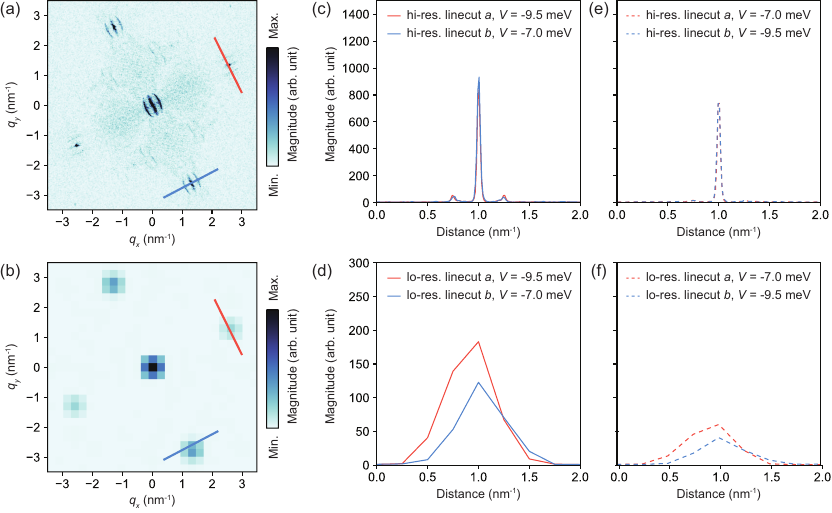}
\caption{\label{fig:4}
Correspondence between symmetry-breaking modulations in Figs. 3 and 4 of the main text. (a) and (b) Fourier transform images from Figs. 3 and 4 of the main text, respectively, shown at the same scale in \textit{q}-space. (c) and (d) Intensity linecuts for each dataset, acquired by sampling across the Bragg peaks as shown by the blue and red lines in (a) and (b), at the energies at which Landau sublevels are present in each valley ($V = -\SI{9.5}{meV}$ and $-\SI{7.0}{meV}$ for valleys \textit{a} and \textit{b}, red and blue, respectively). (e) and (f) Intensity profiles acquired for each at the energies where Landau sublevels are absent in the sampled valley, for comparison with (c) and (d).
}
\end{figure*}

In the main text Figs. 3 and 4 show images extracted from conductance mapping measurements in a large ($80 \times \SI{80}{nm^{2}}$) and small ($4 \times \SI{4}{nm^{2}}$) field-of-view, respectively. The former allows a high \textit{q}-space resolution in Fourier transform images, where `bracket' QPI signals appear near the Bragg peaks, but these are un-resolved in the latter measurement which has a much lower \textit{q}-space resolution. We argue that the symmetry breaking seen in Bragg peak intensity in Fig. 4 corresponds to a contribution from the symmetry breaking QPI in Fig. 3. We discuss this in more detail here. Fig. S4 shows Fourier transform images from the two measurements at the same scale. The structures around the Bragg peaks that are clear for the $80 \times \SI{80}{nm^{2}}$ measurement [panel (a)] are below the resolution limit for the $4 \times \SI{4}{nm^{2}}$ measurement [panel (b)]. Figure S4 also shows that the difference in intensity of the Bragg peaks in the high-resolution data for Fig. 3 [see panels (c) and (e)], as well as in the data for Fig. 4 [see panels (d) and (f)].

The smaller field-of-view where the measurement for Fig. 4 was acquired was located reasonably far from any scattering impurities. Despite this, it does detect QPI from remote scattering centers. The $L$($\mathbf{r}$) images of Fig. 3 clearly show that QPI modulations cover the whole surface, and there are effectively no locations where a narrowing of the field-of-view would allow us to fully avoid them. Therefore, it is very likely that QPI has at least some role in the asymmetry seen in Fig. 4. 

As an aside, the reason that the QPI modulations cover the whole surface is probably a very long coherence length (several tens of nanometers) at low energy, which may be understood by considering the energy-time manifestation of the Heisenberg uncertainty relation as outlined recently by He \textit{et al.} \cite{He2021}. The effect of apparently diverging coherence length near $E_{\mathrm{F}}$ is not unique to ZrSiS and is observed in other circumstances such as at the surface of Cu(111).

\begin{figure*}
\centering
\includegraphics[scale=1]{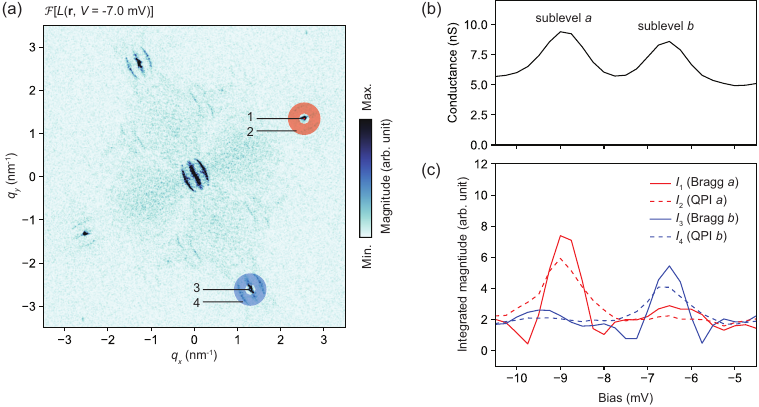}
\caption{\label{fig:5}
Intensity-energy plots discriminating the Bragg peaks and QPI signals in the data for Fig. 3 of the main text. (a) Example $\mathcal{F} \left[ L(\mathbf{r}, V \right] $ image showing the mask locations used to integrate the Bragg peak intensities $I_{1}$ and $I_{3}$, and for the QPI intensities $I_{2}$ and $I_{4}$. The masks for selecting the QPI signals exclude the Bragg peaks. (b) Average of d$I$/d$V$($V$) over the whole \textit{r}-space field of view shown in Fig. 3(a) and (b) of the main text. (c) $I_{1,2,3,4}$($V$) curves.
}
\end{figure*}

To further elucidate the relation between the Bragg peak and QPI intensities, we have also used intensity-vs-energy plots, and to discriminate between Bragg peaks and QPI using suitable masks in Fourier space. This is done as shown in Fig. S5 below. The curves $I_{1(3)}$($V$) are the intensities for the Bragg peaks and $I_{2(4)}$($V$) are the QPI intensities for valley \textit{a}(\textit{b}). Note that the masks for integrating the intensity of QPI signals safely exclude any contribution from the Bragg peaks.

From Fig. S5 we can see that the QPI and Bragg peaks intensities each show a clear asymmetry of their own, but follow each other quite closely. A reasonable interpretation is that the QPI pattern includes scattering with a $q$ vector that coincides with the Bragg vector $\mathbf{G}$, and this is why the data in Fig. 4 has the appearance that it does. The key point is that the symmetry-breaking modulations in Fig. 4 are neatly explained as originating from QPI, without invoking a separate phenomenon such as orbital ordering.

\clearpage

\section*{Supplementary Note 5 - Intra-unit-cell sampling and fitting of d$I$/d$V$ spectra} 

Figure S6 below shows measured LL spectra at various magnetic fields. The tunneling conductance measurements were performed in the same field-of-view as shown in Fig. 4 of the main text. Panel (a) shows the spatially averaged conductance, and panel (b) shows the valley-disaggregated curves obtained using the same procedure as is described in Fig. 4 and the accompanying text. The energies of all the peaks in each valley's LL spectrum must be extracted using curve fitting. Because of the large number of similar measurements in this series, we process the raw data for each field using a script that automates the following workflow:

\begin{itemize}

  \item Flatten the simultaneously acquired topography for the given d$I$/d$V$($\mathbf{r}$, $V$) measurement and find the reciprocal lattice vectors of the atomic corrugation using 2D Gaussian fitting to the Bragg peaks its Fourier transform.
  
  \item Use these reciprocal lattice vectors to generate an $r$-space lattice of sampling points. The bridge site (and hollow site) lattices are generated by shifting the initial atom-centered lattice by half of the measured lattice constant along either of the lattice vectors (and along both lattice vectors for the hollow sites).

  \item Sample the d$I$/d$V$($\mathbf{r}$, $V$) data at all the points obtained as above, and compute the averaged and valley-disaggregated curves.

\end{itemize}

\begin{figure*}
\centering
\includegraphics[scale=1]{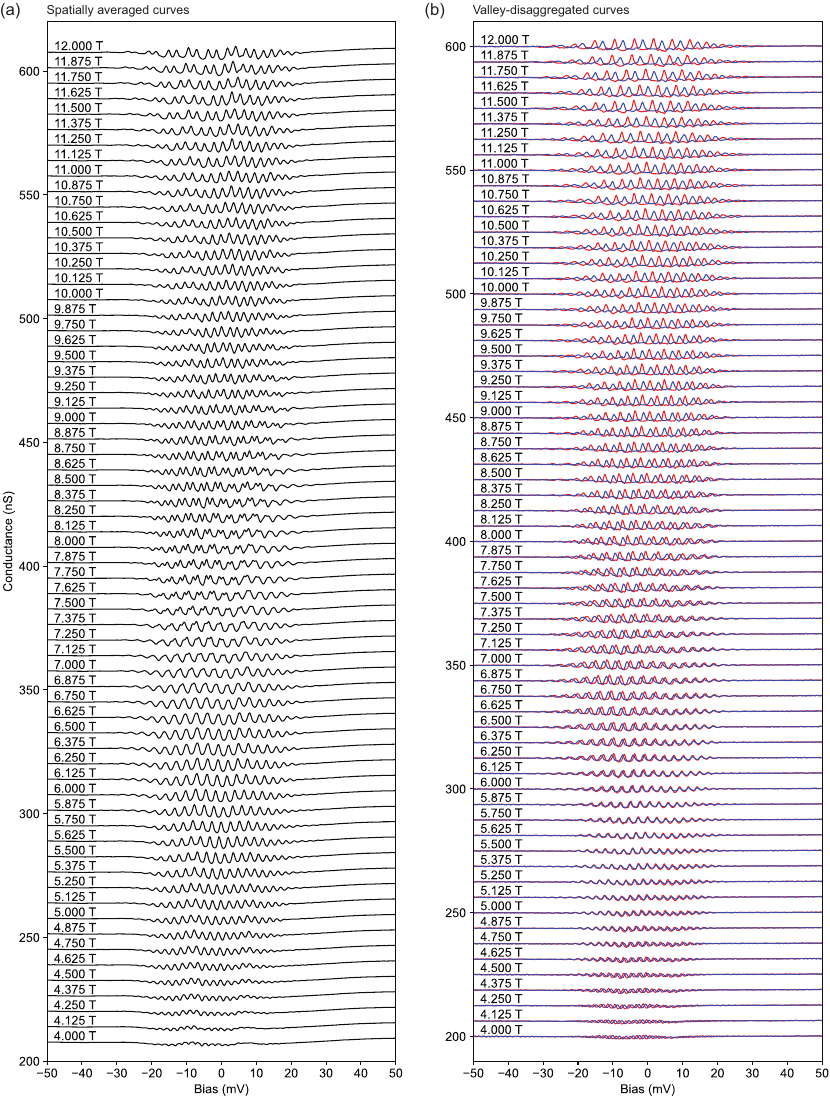}
\caption{\label{fig:6}
Valley-disaggregation of d$I$/d$V$($V$) curves acquired under varying magnetic field. (a) Averages of the measured curves over the $4 \times \SI{4}{nm^{2}}$ field-of-view shown in Fig. 4 of the main text ($V = \SI{50}{mV}$, $I_{\mathrm{set}} = \SI{500}{pA}$, $V_{\mathrm{mod.}} = \SI{0.25}{mV}$). (b) Valley-selective curves after implementing the intra-unit-cell sampling procedure described for Fig. 4 of the main text.}
\end{figure*}

This implements the procedure depicted in Fig. 4(f) and 4(g) of the main text, for the data measured at each field.

Having obtained the curves shown in Fig. S6(b), another script is used to perform a fitting procedure to estimate the energies and other properties of the peaks in each curve. The main challenge is that the number of LLs expected to appear in each curve varies, so for each curve a new fitting model with a suitable number of peaks needs to be constructed on-the-fly and without any human intervention. Figure S7 shows how this is achieved.

The model used for fitting is composed of a Gaussian lineshape whose prefactor $A_{\mathrm{back.}}$ is negative, which describes a background on which the LLs lie. The LLs themselves are each described by a Lorentzian lineshape. Accordingly, the conductance curve is modeled as

\begin{align}
    \frac{\mathrm{d}I}{\mathrm{d}V}(eV) \approx &\frac{A_{\mathrm{back.}}}{\sigma_{\mathrm{back.}} \sqrt{2 \pi}} \exp{\left[- \frac{(eV - E_{\mathrm{back.}})^{2}}{2 \sigma ^{2}} \right] } \\
    &+ \sum_{i}^{N} \frac{A_{i}}{\pi} \frac{\Gamma_{i}/2}{(eV - E_{i})^{2} + (\Gamma_{i}/2)^{2}}. \notag
\end{align}

\begin{figure}[b]
\centering
\includegraphics[scale=1]{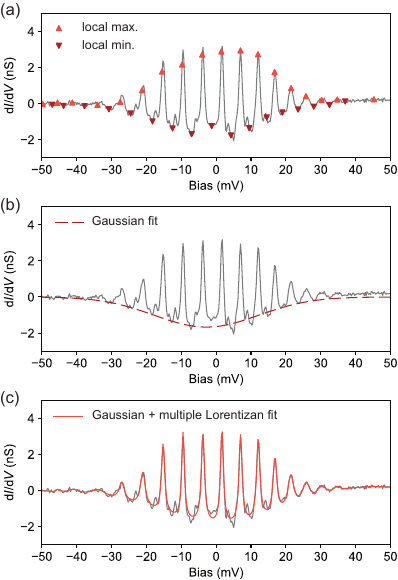}
\caption{\label{fig:7}
Fitting procedure for estimating LL energies. (a) One of the valley-selective d$I$/d$V$($V$) curves obtained using the sublattice sampling method described in Fig. 4 of the main text (specifically, the red curve for $B = \SI{12}{T}$ shown in Fig. S6 above). The local maxima and minima are identified after a 20 point Gaussian smoothing (the curve has 401 points), and these are marked with pink and red triangles, respectively (on top of the \textit{non}-smoothed data). (b) In an intermediate step, a background function is obtained from the set of local minima using fitting with a Gaussian function. (c) Fitting to the non-smoothed data using the full model described by Eq. S1. (The parameters for the Gaussian background are still free parameters of the model.)
}
\end{figure}

\begin{figure*}
\centering
\includegraphics[scale=1]{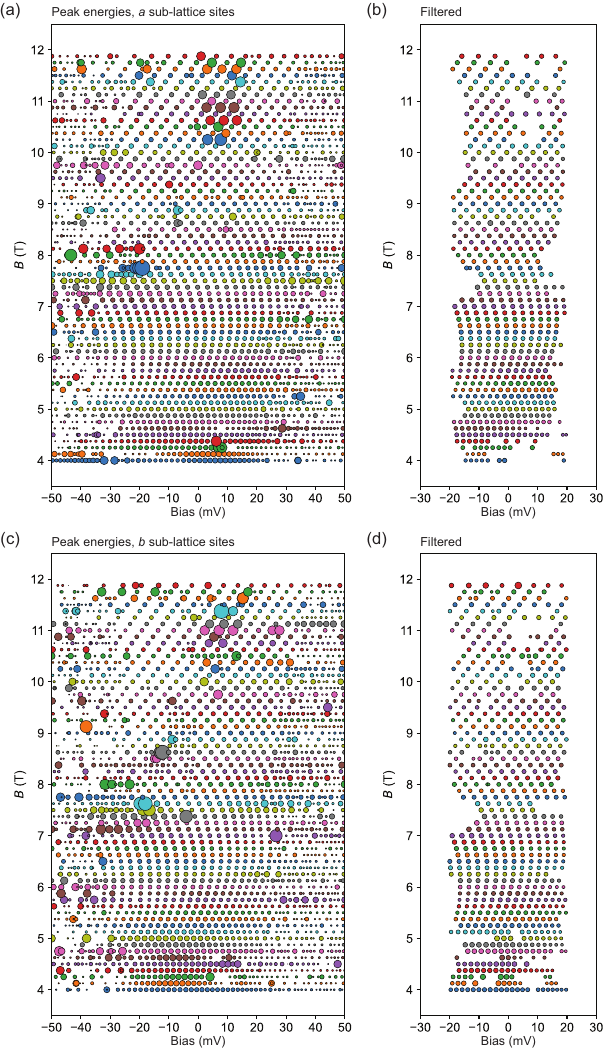}
\caption{\label{fig:8}
Filtering of peak fitting results. (a) Energies of peaks found in d$I$/d$V$ curves sampled at the $a$ sites at all magnetic fields (red curves in Fig. S6). A large number of spurious peaks, for example caused by noise, are found. (b) The remaining points after filtering according to the criteria described in the text. (c) and (d) The corresponding data points obtained from sampling at the $b$ sites (blue curves in Fig. S6).
}
\end{figure*}

After some smoothing of the raw d$I$/d$V$($V$) curves (a 20 point Gaussian convolution -- each curve has 401 points), finding the local maxima and minima of a curve gives a first guess at the locations of each of the peaks. In an intermediate step, the Gaussian background is fitted to the local minima, but the parameters for the Gaussian are left as free parameters for the final fitting step. Finally, multiple Lorentzian lineshapes (one for each local maximum) are included to represent peaks in the spectrum. The initial guess of the energy for each peak is informed by the position of a local maximum, and the initial guess for the intensity is informed by the signal value at that location. The initial guess for all peak widths is $\Gamma_{i,\mathrm{ini.}} = \SI{1}{meV}$.

In practice these peaks include both LLs and noise, and the initial step which picks out the local maxima cannot distinguish between them. We proceed with the fitting anyway and use post-fitting filtering of the data points to exclude noise.

We exclude data points that fall outside of a chosen energy interval and those with un-physical features such as a negative intensity or a width smaller than the experimental resolution. We also impose a reasonable cut-off value for the intensity $A_{i}$. Thus the filtering criteria are:

\begin{itemize}

  \item $\SI{-20}{meV} < E_{i} < \SI{20}{meV}$

  \item $\Gamma_{i} > 3.5 k_{\mathrm{B}}T \approx \SI{0.45}{meV}$
  
  \item $0 < A_{i} < \SI{5}{nS}$

\end{itemize}

The result is that almost all the remaining peaks correspond to LLs, but many peaks that correspond to LLs are excluded (see Fig. S8) and there is some room for improvement.

\section*{Supplementary Note 6 - Landau level peak widths} 

From the above fitting procedure we obtain not only the LL energies but also their widths, as characterized by the broadening parameter $\Gamma_{i}$ of the Lorentzian lineshape fitted to each peak. In Fig. S9(a) we show a plot of values $\Gamma_{i}$ for the $i^{\mathrm{th}}$ peak before filtering out spurious points. Beyond $\sim \SI{20}{meV}$ from $E_{\mathrm{F}}$ the widths of the peak are seen to abruptly increase, which would be the expected result of a damping effect. However there is only a low confidence that these point actually correspond to LLs. In Fig. S9(b) we show the remaining points after the filtering described above. Here we see that the average width is $\Gamma_{i} \sim \SI{1}{meV}$, and there is no strong and consistent relation between the bias voltage and $\Gamma_{i}$. Figure S9(c), which displays the average value of $\Gamma_{i}$ with varying magnetic field, shows no systematic field dependence.

\begin{figure*}[t]
\centering
\includegraphics[scale=1]{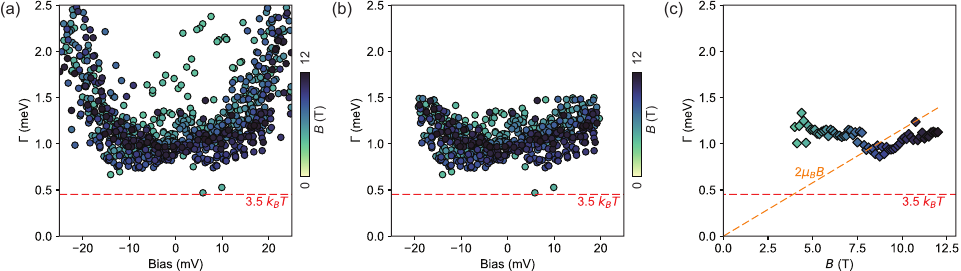}
\caption{\label{fig:9}
Landau level peak widths. (a) Peak widths $\Gamma_{i}$, for the peaks found in the curves measured at the $a$ bridge sites and for all magnetic fields. The energy corresponding to the experimental energy resolution, $3.5 k_{\mathrm{B}}T$ \cite{ChenBook} with $T= \SI{1.5}{K}$, is marked with a pink dashed line. (b) The widths of the peaks remaining after the filtering described for Fig. S8. (c) The average value of the peak widths for each magnetic field. The yellow dashed line gives the Zeeman energy for a free electron as a function of magnetic field.
}
\end{figure*}

In each panel we also show the energy corresponding to our experimental resolution, showing that the apparent width of the LLs is intrinsic and not simply the result of experimental limitations. (Additional measurements performed at $T = \SI{0.35}{K}$ confirm this.) The intrinsic width of the Landau levels is attributable to quasiparticle scattering. Invoking the energy-time uncertainty relation $\Delta E \cdot \Delta t = \Gamma \cdot \tau \geq \hbar / 2$, the width corresponds to a timescale of $\tau = \SI{6.58 e-13}{s}$. Given the measured average velocity of the surface band, $\tilde{v} \approx  \SI{4.5 e5}{m/s}$, this would correspond to a quasiparticle mean free path of about $\SI{150}{nm}$. Another way to characterize the scattering rate for a Landau level is as a Dingle temperature $T_{\mathrm{D}} = \hbar / \left( 2 \pi k_{\mathrm{B}} \tau \right) \approx \SI{1.85}{K}$ \cite{Dingle1952}.

\section*{Supplementary Note 7 - Spatial mapping of strain} 

As well as significant sample dependence, the valley splitting also shows some spatial variation. In the measurements shown for sample C8 in the main text, this can be characterized by the variation in the energy difference between the two valley-polarized sublevels visualized in Fig. 3. The d$I$/d$V$($\mathbf{r}$, $V$) measurement from which the images in Fig. 3 are derived spans the energy range from $\SI{-10.5}{}$ to $\SI{-4.5}{meV}$, covering two peaks. We fit a function to these peaks consisting of a pair of Lorentzian lineshapes on a linear background:

\begin{align}
    \frac{\mathrm{d}I}{\mathrm{d}V}(eV) \: \approx \ &\frac{A_{a}}{\pi} \frac{\Gamma_{a}/2}{(eV - E_{a})^{2} + (\Gamma_{a}/2)^{2}} \\
    &+ \frac{A_{b}}{\pi} \frac{\Gamma_{b}/2}{(eV - E_{b})^{2} + (\Gamma_{b}/2)^{2}} \notag \\
    &+ m(eV) + c  \notag
\end{align}

\begin{figure}
\centering
\includegraphics[scale=1]{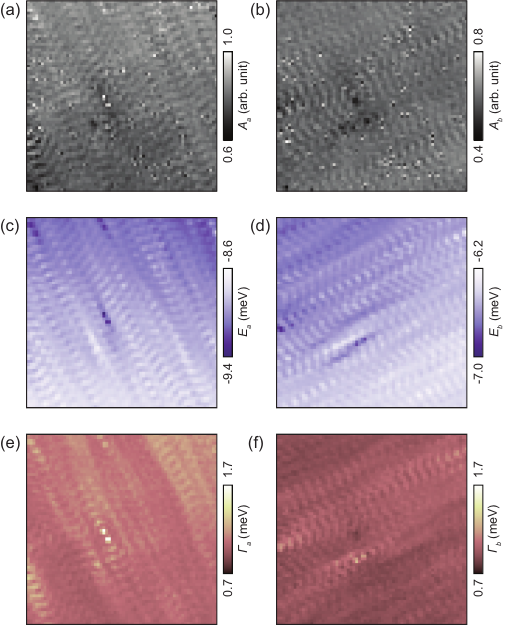}
\caption{\label{fig:10}
Results of fitting a pair of Lorentzian functions to the pair of LL peaks in the d$I$/d$V$($\mathbf{r}$, $V$) data underpinning Fig. 3 of the main text. Fitting is performed at each pixel after a 10$\times$10 pixel course-graining of the raw data. (a) Intensities $A_{a}$ and $A_{b}$. (b) Energies $E_{a}$ and $E_{b}$. (c) Peak widths $\Gamma_{a}$ and $\Gamma_{b}$.
}
\end{figure}

\begin{figure}
\centering
\includegraphics[scale=1]{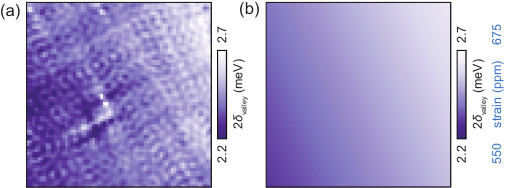}
\caption{\label{fig:11}
Spatial imaging of $2 \delta_{\mathrm{valley}}$. (a) An image of $E_{b} - E_{a} = 2 \delta_{\mathrm{valley}}$. (b) A planar fit to the image in (a). The equivalent range of values for strain are shown alongside the colorbar.
}
\end{figure}

Figure S10 shows the resulting optimized fitting parameters for $A_{a(b)}$, $E_{a(b)}$ and $\Gamma_{a(b)}$. We then obtain $E_{b} - E_{a} = 2 \delta_{\mathrm{valley}}$, which is plotted in Fig. S11(a). Although the map of $2 \delta_{\mathrm{valley}}$ shows spatial modulations that resemble the quasiparticle interference modulations in Fig. 3 of the main text, these are probably a spurious artifact of the fitting procedure. The global trend, clarified by the planar fit shown in Fig. S11(b), is a finite variation of $2 \delta_{\mathrm{valley}}$ over larger length scales. The corresponding values for local strain, that we estimate using arguments given in the main text, are also shown. The field of view is $80 \times \SI{80}{nm^{2}}$, and we see a variation of $\SI{125}{ppm}$ over a $\SI{100}{nm}$ length scale. Extrapolating this variation out to a scale approaching the size of the sample ($\sim \SI{1}{mm}$) quickly results in unfeasibly large strain. The actual distribution must have significant inhomogeneities, e.g. ripples or wrinkles, on scales larger than the field-of-view shown here, but much smaller than the sample size.

\section*{Supplementary Note 8 - Considerations for LL indices \& Berry phase search procedure} 

To determine the LL indices and Berry phase around the surface band contours, we use a process of minimizing the spread, characterized as $\sigma$, of the LL energy points in a $E$-versus-$\nu B$ plot (see Fig. 5 of the main text). Here we investigate how the effectiveness of this process varies as fewer data points are used, representing a shorter measurement. First, we select the subset for $B \le \SI{9}{T}$ (for a more economical superconducting magnet configuration, solid blue curve in Fig. S12). We also try subsets of the data at increasingly sparse values of magnetic field strength. Instead of using an increment of $\Delta B = \SI{0.125}{T}$, as shown in the main text, we take only the subset of data at increments of $\Delta B = \SI{0.25}{T}$, $\SI{0.5}{T}$, and $\SI{1}{T}$ (dashed curves in Fig. S12).
In all cases $\sigma$ is minimized closer to $\nu_{\mathrm{offset}} = 24.5$ than to 24 or 25, in agreement with the result shown in the main text and indicating a Berry phase of $\gamma_{B} = 0$. The result is fairly convincing if an increment of $\Delta B \le \SI{0.5}{T}$ is used, with 154 data points remaining, but appears less reliable for greater increments and fewer points.

Given that an increment of $\Delta B = \SI{0.5}{T}$ would lead to a decisive result, the challenge would then be the assignment of provisional indices $\nu_{\mathrm{prov.}}$. This is because if $\nu_{\mathrm{prov.}}$ is assigned by hand, it is difficult to track which peaks in different conductance curves belong to the same LL if the change in field causes the LL energy to change by more than a fraction of the energy interval between two LLs. We leave this as an issue to be addressed in the future.

\begin{figure}
\centering
\includegraphics[scale=1]{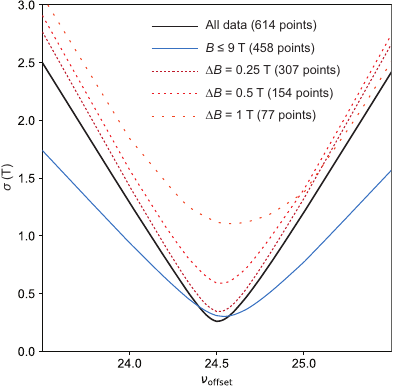}
\caption{\label{fig:12}
Minimization of the spread, $\sigma$, of LL energy data points in $E$-versus-$\nu B$ space, using only various subsets of the total data set.
}
\end{figure}

From Eqn. 4 of the main text, while $\sigma$ is the average contribution to the spread of the data points in a $E$-versus-$\nu B$ plot, $\sigma N$ is the total spread. The minimum possible value for $\sigma N$ is the span of the points along the $\nu B$ axis, but this simply corresponds to the span along the $E$ axis where LLs can be observed [see Fig. 5(b) of the main text as an example], which depends on such details as the width of the band hosting LLs and, as we speculate in the main text, attenuation of the LL peaks due to \textit{e}-\textit{ph} interactions.